\documentclass[10pt]{article}
\usepackage[final]{epsfig}
\usepackage{amsfonts, amsmath, amssymb, amsthm}
\usepackage{latexsym}
\usepackage{color}

\numberwithin{equation}{section}

\newtheorem{theorem}{Theorem}[section]
\newtheorem{lemma}[theorem]{Lemma}

\newtheorem{example}[theorem]{Example}
\newtheorem{proposition}[theorem]{Proposition}
\newtheorem{definition}[theorem]{Definition}
\newtheorem{corollary}[theorem]{Corollary}

\newcommand{\R}{\mathbb{R}}

\newcommand{\C}{\mathbb{C}}
\newcommand{\Z}{\mathbb{Z}}
\newcommand{\N}{\mathbb{N}}

\newcommand{\be}{\begin{equation}}
\newcommand{\ee}{\end{equation}}

\newcommand{\cala}{{\mathcal{A}}}
\newcommand{\calb}{{\mathcal{B}}}
\newcommand{\calR}{{\mathcal{R}}}
\newcommand{\calP}{{\mathcal{P}}}

\newcommand{\bthe}{\begin{theorem}}
\newcommand{\ethe}{\end{theorem}}

\newcommand{\ben}{\begin{enumerate}}
\newcommand{\een}{\end{enumerate}}

\newcommand{\beq}{\begin{equation}}
\newcommand{\eeq}{\end{equation}}

\newcommand{\ble}{\begin{lemma}}
\newcommand{\ele}{\end{lemma}}

\newcommand{\bde}{\begin{definition}}
\newcommand{\ede}{\end{definition}}

\newcommand{\bco}{\begin{corollary}}
\newcommand{\eco}{\end{corollary}}

\newcommand{\bpr}{\begin{proposition}}
\newcommand{\epr}{\end{proposition}}

\newcommand{\bexam}{\begin{example}\rm}
\newcommand{\eexam}{\halmos\end{example}}

\newcommand{\beao}{\begin{eqnarray*}}
\newcommand{\eeao}{\end{eqnarray*}\noindent}

\newcommand{\beam}{\begin{eqnarray}}
\newcommand{\eeam}{\end{eqnarray}\noindent}

\newcommand{\barr}{\begin{array}}
\newcommand{\earr}{\end{array}}

\newcommand{\bproof}{\begin{proof}}
\newcommand{\eproof}{\end{proof}}

\newcommand{\eps}{\varepsilon}

\newcommand{\xhat}{\widehat{x}}

\def\calR{{\mathcal{R}}}

\newcommand{\halmos}{\quad\hfill\mbox{$\Box$}}

\begin{document}

\begin{center}
{\LARGE Smoothing of transport plans with fixed marginals and rigorous semiclassical limit of the Hohenberg-Kohn functional}  \\

\normalsize
\vspace{0.4in}

Codina Cotar$^1$, Gero Friesecke$^2$ and Claudia Kl\"uppelberg$^2$  \\[1mm]
$^1\,$Department of Statistical Science, University College London \\
$^2\,$Faculty of Mathematics, Technische Universit\"at M\"unchen  \\
c.cotar@ucl.ac.uk, gf@ma.tum.de, cklu@ma.tum.de \\[2mm]
\end{center}

\noindent
\normalsize
{\bf Abstract.} We prove rigorously that the exact N-electron Hohenberg-Kohn density functional converges in the strongly interacting limit to the strictly correlated electrons (SCE) functional, and that the absolute value squared of the associated constrained-search wavefunction tends weakly in the sense of probability measures to a minimizer of the multi-marginal optimal transport problem with Coulomb cost associated to the SCE functional. This extends our previous work for $N=2$ \cite{CFK11}. The correct limit problem has been derived in the physics literature by Seidl \cite{Se99} and Seidl, Gori-Giorgi and Savin \cite{SGS07}; in these papers the lack of a rigorous proof was pointed out.

We also give a mathematical counterexample to this type of result, by replacing the constraint of given one-body density -- an infinite-dimensional quadratic expression in the wavefunction -- by an infinite-dimensional quadratic expression in the wavefunction and its gradient. Connections with the Lawrentiev phenomenon in the calculus of variations are indicated. 

\vspace{0.2in}




\vspace{0.2in}

\large

\section{Introduction}
During the past decades, density functional theory (DFT) has become the standard method for numerical electronic structure computations in physics, chemistry and materials science, due to its combination of low computational cost yet surprisingly high accuracy. DFT approximately recovers the quantum mechanical ground state energy of (high-dimensional) many-electron systems via variational principles for the (low-dimensional) one-body density. See \cite{HK64, KS65} for original papers, \cite{PY95} for a classic textbook account, and \cite{Be14} for a recent review. 

This paper aims to make a mathematical contribution to DFT. We prove rigorously that the ``exact'' Hohenberg-Kohn (HK) density functional converges in the semiclassical limit to the strictly correlated electrons (SCE) functional. In the original papers in the physics literature in which the limit functional was derived \cite{Se99, SGS07}, the authors Seidl, Gori-Giorgi and Savin pointed out that ``there is no rigorous proof for this reasonable conjecture'' (\cite{SGS07}, p. 042511-2). 

We postpone further discussion for the moment and first describe the above functionals in more detail. 

The variational problem underlying the SCE functional is the following optimal transport problem: 
\be \label{VP1}
   \mbox{Minimize }\tilde{V}_{ee}[\gamma] = \int_{\R^{3N}} \sum_{1\le i<j\le N} 
   \frac{1}{|x_i-x_j|} \, d\gamma(x_1,..,x_N) \mbox{ over }\gamma\in\calP_{sym}(\R^{3N}) 
   \mbox{ subject to }\gamma\mapsto\rho/N.
\ee
Here $\calP_{sym}(\R^{3N})$ denotes the set of symmetric probability measures on $\R^{3N}$, where symmetric means
$$
   \gamma(A_1\times\cdots\times A_N) = \gamma(A_{\sigma(1)}\times\cdots\times A_{\sigma(N)}) \mbox{ for all measurable sets } A_1,..,A_N\subseteq\R^3 
\mbox{ and all permutations } \sigma;
$$
$\rho$ is a given nonnegative integrable function on $\R^3$ with $\int_{\R^3}\rho = N$ (physically: the total electron density of an atom or molecule with $N$ electrons);
and the notation $\gamma\mapsto\rho$ means that $\gamma$ has one-body density $\rho$ (physics language) alias $\R^3$-marginal density $\frac1N\rho$ (probability language), i.e.
\be \label{marginals}
   \gamma(A\times\R^{3(N-1)}) = \int_A \frac{\rho(x)}{N} \, dx \mbox{ for all measurable }
   A\subseteq\R^3.
\ee
If $\gamma$ is given by an integrable function, i.e. $\gamma = \rho_N \, dx_1..dx_N$ for some integrable $\rho_N$, the above definition reduces to $\rho(x_1) = N \int_{\R^{3N-3}}\rho_N(x_1,..,x_N)\, dx_2..dx_N$. 
Physically, $\gamma$ is the joint probability measure of the positions of $N$ electrons in $\R^3$. The normalization factor $N$ is owed to the convention in quantum mechanics that the one-body density $\rho$ integrates to the number of particles in the system, whereas the marginal density $\mu$ in the sense of probability theory integrates to $1$, so that $\mu=\rho/N$. The SCE functional is the optimal cost in \eqref{VP1} as a function of the one-body density,  
\be \label{FOTKa} 
   V^{SCE}_{ee}[\rho] = \inf_{\gamma\in \calP_{sym}(\R^{3N}), \, \gamma\mapsto \rho/N} 
   \tilde{V}_{ee}[\gamma].
\ee
Mathematically, \eqref{VP1} is a multi-marginal optimal transport problem; note that, on account of the symmetry of $\gamma$, condition (\ref{marginals}) means
that all the $N$ $\R^3$-marginals of the N-body density $\gamma$ are prescribed. Moreover the cost is a Coulomb cost -- not the usual positive power of the distance but a negative power. 

The fact that \eqref{VP1} can be interpreted as an optimal transport problem, and its rigorous formulation via probability measures above, is due to \cite{CFK11, BDG12}, and has led to great interest in the mathematics literature as the repulsive nature of the cost induces new phenomena and challenges (see \cite{DGN15} for a survey and references). 
A particularly interesting question which we briefly review here is whether minimizers are of Monge form
\be \label{Monge}
   \gamma(x_1,..,x_N) = \frac{\rho(x_1)}{N} \delta_{T_2(x_1)}(x_2) \cdots \delta_{T_N(x_1)}(x_N)
\ee 
(or a symmetrization thereof) for some transport maps $T_2,..,T_N:\R^d\to\R^d$. This low-dimensional form, when taken together with the convergence of the HK functional to \eqref{FOTKa}, would show that the {\it curse of dimensionality disappears from the HK functional in the semiclassical limit, despite the limit being strongly correlated.} The form \eqref{Monge} was suggested in the original papers \cite{Se99, SGS07}, and explains the name ``strictly correlated electrons'': while the position of the first electron possesses a spread-out probability distribution, given this position -- say, $x_1$ -- the positions of all the other electrons are deterministic, and in particular strongly correlated. 

As turns out, however, {\it non}-Monge minimizers of \eqref{VP1} exist when $N\ge 3$, as was discovered by Pass \cite{Pa13}. This is in striking contrast to the standard situation of two-marginal problems with positive-power cost \cite{Vi09}, and is due to the {\it combined} effects of (i) more than two marginals, (ii) negative-power costs, and (iii) single-particle space dimension bigger than one. If any one of these three complications is dropped, minimizers have been proven to be unique and of Monge form: when either $N=2$ \cite{CFK11,BDG12}, or the cost is a positive-power cost such as the harmonic cost \cite{GS98} (see \cite{He02, Ca03, KP14} for generalizations), or when the single-particle space dimension is one \cite{CDD13, MP17}. Whether Monge minimizers exist for \eqref{VP1} is open when $N\ge 3$; moreover the minimizer is unique and non-Monge when $N=\infty$ \cite{CFP15}.
\\[2mm]
Next we describe the Hohenberg-Kohn functional. It is defined as follows 
\beq \label{FHK}
   F_{\alpha}^{HK}[\rho] := \inf_{\Psi\in\cala_N, \, \Psi\mapsto\rho} \Big\{\alpha T[\Psi]+V_{ee}[\Psi]\Big\},
\eeq
where $\alpha>0$ is a semiclassical coupling constant whose physical meaning is explained below, and the admissible class of $N$-particle wavefunctions is
\beq \label{class}
  \cala_N = \{ \Psi \in L^2((\R^3\times\Z_2)^N;\C) \, | \, \nabla\Psi \in L^2, \, \Psi \,  \mbox{antisymmetric}, \,
                ||\Psi||_{L^2}=1\},
\eeq
where antisymmetric means 
\be \label{anti}
    \Psi(z_{\sigma(1)} .., z_{\sigma(N)}) = sgn(\sigma)\Psi(z_1,..,z_N) \mbox{ for all permutations }\sigma
\ee
(with $z_1,...,z_N\in\R^3\times\Z_2$ denoting space-spin-coordinates for the $N$ electrons). This implies in particular that the corresponding $N$-body density of electron positions,
\be \label{posden}
   \rho_N^\Psi(x_1,..,x_N) = \sum_{s_1,..,s_N\in\Z_2} |\Psi(x_1,s_1,..,x_N,s_N)|^2,
\ee
is symmetric. The standard notation $\Psi\mapsto\rho$ means that the one-body density of $\Psi$, 
\beq \label{density}
   \rho^{\Psi}(x_1) = N \int_{\R^{3(N-1)}} \sum_{s_1,..,s_N\in\Z_2} |\Psi(x_1,s_1,..,s_N,x_N)|^2 dx_2..dx_N, 
\eeq
is equal to $\rho$. The associated class of one-body densities
on which the HK functional \eqref{FHK} is defined is the image of $\cala_N$ under the map (\ref{density}),
\be
    \calR_N = \{\rho^\Psi \, | \, \Psi\in\cala_N\}.
\ee
By a result of Lieb \cite{Li83}, it equals 
\beq \label{densityclass}
   \calR_N = \{ \rho \, : \, \R^3 \to \R \, | \, \rho\ge 0, \, \sqrt{\rho}\in H^1(\R^3), \;
   \int_{\R^3}\rho(x)\, dx = N\},
\eeq
where $H^1(\R^3)$ is the usual Sobolev space $\{ u \in L^2(\R^3) \, | \, \nabla u\in L^2(\R^3)\}$. Finally, the functionals $T$ and $V_{ee}$ are 
\beq \label{T}
  T[\Psi] = \int_{\R^{3N}} \sum_{s_1,..,s_N\in\Z_2} |\nabla_{x_i}\Psi(x_1,s_1,..,x_N,s_N)|^2 dx_1..dx_N
\eeq
(kinetic energy),
\beq \label{Vee}
  V_{ee}[\Psi] =  \int_{\R^{3N}} \sum_{s_1,..,s_N\in\Z_2} \sum_{1\le i < j \le N} \frac{1}{|x_i-x_j|} |\Psi(x_1,s_1,..,x_N,s_N)|^2 dx_1..dx_N
\eeq
(electron-electron interaction energy). Here we have used atomic units
$\hbar=m=|e|=1$, where $m$ and $e$ are mass and charge of the electron and $\hbar$ is the reduced Planck constant. For integral of the form \eqref{Vee}, i.e. $\int_{\R^{3N}}f(x_1,..,x_N) \sum_{s_1,..,s_N\in\Z_2}|\Psi(x_1,s_1,..,x_N,s_N)|^2 dx_1..dx_N$, we will often use the shorthand $\int_{\R^{3N}} f \sum_{s_1,..,s_N\in\Z_2}|\Psi|^2$. 

The above variational definition of the HK functional, and the underlying ``constrained-search'' problem 
\be \label{VP2}
  \mbox{Minimize }\alpha T[\Psi] + V_{ee}[\Psi] \mbox{ over }\Psi\in\cala_N 
  \mbox{ subject to }\Psi\mapsto\rho,
\ee
were introduced in \cite{Le79, Li83}, simplifying the original indirect and non-variational construction \cite{HK64}. Also, we note that minimizers of \eqref{VP2} are known to exist for any $\rho\in\calR_N$ \cite{Li83}. The importance of this functional lies in the fact that it allows to predict the {\it exact} electronic ground state energy $E_0$ of any molecular system, via the formula
$$
  E_0 = \inf_{\rho\in\calR_N} \Bigl(F^{HK}_1[\rho] + \int_{\R^3} v\, \rho\Bigr),
$$
where $v(r)=-\sum_{\alpha=1}^M Z_\alpha |r-R_\alpha|^{-1}$ is the Coulomb potential exerted by the atomic nuclei, with $Z_\alpha>0$ and $R_\alpha\in\R^3$ denoting the nuclear charges and positions. (For a mathematical account see e.g. \cite{Li83} or \cite{CFK11} Sec. 2.)

We can now state our main results precisely.
\\[3mm]
{\bf Theorem A} (Energy) {\it For any $\rho\in\calR_N$,} $\lim\limits_{\alpha\to 0} F^{HK}_\alpha[\rho] = V^{SCE}_{ee}[\rho]$.
\\[3mm]
{\bf Theorem B} (Wavefunction) {\it For any $\rho\in\calR_N$, and any sequence $\{\Psi_{\alpha}\}_{\alpha>0}$ of minimizers of $\alpha T+V_{ee}$ subject to the constraint $\Psi_\alpha\mapsto\rho$, there exists a subsequence such that as $\alpha\to 0$, 
$$
   \sum_{s_1,..,s_N\in\Z_2} \!\!\! |\Psi_\alpha|^2 \; \rightharpoonup \; \gamma
$$
for some minimizer $\gamma\in\calP_{sym}(\R^{3N})$ of the optimal transport problem in \eqref{FOTKa}. Here the halfarrow $\rightharpoonup$ denotes weak convergence of probability measures, that is to say 
\be \label{WF2}
    \lim_{\alpha\to 0} \int_{\R^{3N}} f \!\! \sum_{s_1,..,x_N\in\Z_2}\!\!\!\! |\Psi_\alpha|^2 = \int_{\R^{3N}}\! f \; d\gamma \mbox{ for all bounded continuous functions }f:\R^{3N}\to\R. 
\ee
} 
\\[3mm]
Physically, Theorem B says that the optimal transport problem \eqref{FOTKa} correctly describes not just the asymptotic energy of the quantum system, but the asymptotic behaviour of any observable of form \eqref{WF2}. Mathematically, such a convergence of minimizers can be interpreted as an instance of Gamma convergence, see Theorem \ref{T:WF} in Section \ref{sec:WF} for a precise statement. For careful numerical comparisons of wave functions squared and optimal plans for small $\alpha$ see \cite{CF15}.

These results extend our previous work for $N=2$ \cite{CFK11} to $N$ particles, and were announced in \cite{CFP15}. Similar extensions were recently obtained independently in the preprints \cite{BD17} and \cite{Le17}. In \cite{BD17}, Bindini and De Pascale introduce a somewhat different regularization technique from ours \cite{CFK11} and establish Theorems A and B for $N=3$ and for the bosonic analogue of \eqref{VP2}. In \cite{Le17}, Lewin extends their technique to fermionic mixed states, thereby obtaining the analogue of Theorem A for the mixed-state version, or convexification, of \eqref{VP2}.  

Mathematically, proving Theorems A and B requires to overcome
two difficulties: first, optimizers of the limit problem \eqref{VP1} are singular measures which do not arise as N-body densities of any square-integrable function,
let alone one with finite kinetic energy; but smoothing destroys the marginal constraint. Second, admissible trial functions in the HK functional must be antisymmetric;
but antisymmetrizing a given trial function again destroys the marginal constraint. 

To illustrate the first difficulty, we have designed a counterexample which shares many features of \eqref{VP2} but where the semiclassical limit of the energy is strictly bigger than the infimum of the functional obtained by dropping the kinetic term. Our example, presented in Section \ref{sec:counter}, is related to the Lawrentiev phenomenon.

The first difficulty is overcome by extending our technique introduced in \cite{CFK11} to
re-instate the original marginals after smoothing (while preserving the Sobolev regularity $\sqrt{\gamma}\in H^1$) to the multi-marginal case. To deal with transversal crossings of initial and target marginal, we introduced in \cite{CFK11} the idea of ``strong positivization'', i.e. mixing in a small amount of the mean field plan. The correct $N$-body generalization turns out to be mixing in a small amount of a partial mean field plan, consisting of tensor products of one-body and $(N-1)$-body marginals (and not, for instance, an $N$-fold tensor product of one-body marginals); see Section \ref{sec:bosons}. 
The second difficulty is overcome by a novel technique to re-instate marginals after node insertion. One ingredient in the latter are certain clever explicit antisymmetric representations of a given density due to Harriman \cite{Ha81} and Lieb \cite{Li83}. In fact, it turns out that the Harriman-Lieb construction has a beautiful optimal transport meaning, explained in Appendix A.
\\[2mm]
Let us finally turn to the physical meaning of the coupling constant $\alpha$ in \eqref{VP2} and \eqref{FHK}. While originally introduced as a semi-empirical constant which governs the relative strength of kinetic and repulsion energy, it emerges naturally via considering the following {\it dilute scaling limit} \cite{CF15}. Start from a density $\rho$. Scale it via $\rho_\alpha(x):=\alpha^3 \rho(\alpha x)$. Let $\Psi[\rho_\alpha]$ denote a minimizer of the original ($\alpha=1$) variational problem \eqref{VP2} for the scaled density, i.e. a minimizer of $T[\Psi]+V_{ee}[\Psi]$ subject to $\Psi\mapsto\rho_\alpha$. Finally, scale back, i.e. define  $\alpha^{-3N/2}\Psi[\rho_\alpha](\alpha^{-1}x_1,s_1,..,\alpha^{-1}x_N,s_N)=:\Psi^{(\alpha)}$.
Then \cite{PL85, CF15} $\Psi^{(\alpha)}$ is a minimizer of \eqref{VP2}, and the minimum energy in \eqref{VP2} is related to the original Hohenberg-Kohn energy of the scaled density via
$F_\alpha^{HK}[\rho] = \alpha^{-1}F_1^{HK}[\rho_\alpha]$. Hence Theorems A and B say that in the dilute limit $\alpha\to 0$, 
$$
  \sum_{s_1,..,s_N\in\mathbb{Z}_2} \!\!\!\!\! |\Psi^{(\alpha)}|^2 \rightharpoonup \gamma, \;\;\;\;\;\;\; \frac{1}{\alpha} F_1^{HK}[\rho_\alpha] \to V_{ee}^{SCE}[\rho]. 
$$
Electron densities in practical simulations are insufficiently dilute to directly replace the Hohenberg-Kohn functional (or the electron-interaction part of it) by the optimal Coulomb cost. For instance, the binding curve of the hydrogen molecule turns out to be correct at long range, but far off at equilibrium \cite{CFM14}, in contraposition to conventional density functionals, which are remarkably accurate near equilibrium but poor at long range. Nevertheless it appears that the semiclassical, or strictly correlated, limit studied here -- when combined with traditional DFT functionals based on the opposite, weakly correlated limit -- is a promising ingredient in ongoing research to design new types of hybrid functionals 
(see e.g. \cite{FGSD16}). 

The plan of this paper is as follows. Sections 2 and 3 are devoted to the proof of Theorem A, and form the heart of this paper. In Section 4 we show via standard arguments that Theorem A implies Theorem B. Section 5 discusses our counterexample and the connection to the Lawrentiev phenomenon, and the appendix re-visits the Harriman-Lieb orbital representation of a given density from an optimal transport point of view.

%
%
%
%
%
%
%
%
%
%
%
\section{Proof of the semiclassical limit theorem, I: Bosons} \label{sec:bosons}
We split the proof of Theorem A into two parts, in order to attack separately the two issues of
lack of regularity and absence of antisymmetry in the OT problem (\ref{FOTKa}). 

To this end we introduce the following ''intermediate'' functional between $F^{HK}_\alpha$ and $V_{ee}^{SCE}$:
\be\label{FBos}
     F_\alpha^{Bos}[\rho] := \inf_{\Psi\in\calb
_N, \, \Psi\mapsto\rho} \Big\{\alpha T[\Psi]+V_{ee}[\Psi]\Big\},
\ee
where
\be \label{BN}
     \calb_N := \{\Psi\in L^2(\R^{3N};\C) \, | \, \nabla\Psi\in L^2, \, \Psi\mbox{ symmetric}, \,
                  ||\Psi||_{L^2}=1\}.
\ee
Here symmetric means that 
\be \label{sym}
     \Psi(x_{\sigma(1)},..,x_{\sigma(N)}) = sgn(\sigma) \Psi(x_1,..,x_N) \mbox{ for all permutations }\sigma.
\ee
The difference to the original HK functional \eqref{FHK} is that wavefunctions $\Psi\in\calb_N$ are symmetric and spinless as opposed to antisymmetric and spin-dependent. Physically, such wavefunctions describe a 
system of $N$ bosons. We remark that for any $\rho\in\calR_N$, the class \eqref{BN} of wavefunctions in the
infimization \eqref{FBos} is nonempty, for instance one may take $\Psi=\prod_{i=1}^N\sqrt{\rho(x_i)/N}$. 

For bosonic wave functions $\Psi\in\calb_N$, the $N$-body position density is given by the bosonic analogue of eq. \eqref{posden}, 
\be \label{Nbodydensbos}
   \rho_N(x_1,...,x_N) = |\Psi(x_1,..,x_N)|^2.
\ee

In this section we prove: 
\begin{theorem} \label{T:bosons} For any $\rho\in\calR_N$ (see \eqref{densityclass}),  
$\lim_{\alpha\to 0}F_\alpha^{Bos}[\rho] = V_{ee}^{SCE}[\rho]$.
\end{theorem}
The following strategy was introduced in \cite{CFK11} to establish this result when $N=2$, and can be extended to deal with $N$ bosons. 
\begin{itemize}
\item Start from an optimal plan $\gamma$ with one-body density $\rho$.
\item Smooth it. Note that this modifies the one-body density.
\item Make it strongly positive (see Def. 5.7 in \cite{CFK11}, generalized in Def. \ref{D1} below), by mixing in a small amount of a mean-field plan (or, in the $N$-body case, a ``partial mean field plan'', see \eqref{epsbeta}).
\item Re-instate the original one-body density while preserving the Sobolev regularity $\sqrt{\gamma}\in H^{1}$, hence obtaining a plan with the original one-body density which is the position density \eqref{Nbodydensbos} of a bosonic wavefunction.
\item Pass to the semiclassical limit by suitable error estimates on the three modification steps above (smoothing, achieving strong positivity, re-instating the marginal constraint). 
\end{itemize}
The main technical work in the $N$-boson case goes into showing that the natural generalization of our construction in \cite{CFK11} to re-instate the marginal constraint preserves the Sobolev regularity $\sqrt{\gamma}\in H^{1}$. Our proof of this (see Theorem \ref{T:Sobpres} below) requires that the notion of strong positivity introduced for $N=2$ in \cite{CFK11} be extended to $N$-body problems as follows.
\\[2mm]
{\bf Notation.} Here and below we use the following notation which is common in analysis but not in probability theory. Whenever a transport plan $\gamma\in\calP(\R^{Nd})$ has a density $\rho_N$ with respect to the Lebesgue measure, i.e. $\gamma = \rho_N \, dx_1 .. dx_N$, we denote the density $\rho_N$ again by $\gamma$, and write 
$\gamma\in\calP(\R^{Nd})\cap L^1(\R^{Nd})$. 
\begin{definition} \label{D1} A transportation plan $\gamma \in \calP(\R^{Nd})\cap L^1(\R^{Nd})$ is called \emph{strongly positive} if there exists a constant $\beta>0$ such that 
\be \label{SPdef}
  \gamma(x_1,..,x_N) \ge \beta \int \gamma(x_1,..,x_N) \,  d\xhat_i \; \int \gamma(x_1,..,x_N) \,  dx_i \;\; \mbox{ for all }i=1,..,N,
\ee
where $\widehat{x}_i=(x_1,..,x_{i-1},x_{i+1},..,x_N)$ denotes the coordinates other than $x_i$.  
\end{definition}

We note that for $N=2$, \eqref{SPdef} reduces to the condition in \cite{CFK11} that
$\gamma\ge \beta \mu\otimes\nu$, where $\mu$ and $\nu$ are the marginals of $\gamma$; i.e., $\gamma$ must be bounded from below by the associated mean field plan.

\subsection{Re-instating the marginal constraint while preserving Sobolev regularity}
We generalize our construction from \cite{CFK11} to ``re-instate the constraint'', i.e. to deform a given trial plan into a nearby one with prescribed marginals, to $N$-body problems. 

We suppose that we are given a transportation plan $\gamma_A\in\calP(\R^{Nd})\cap L^1(\R^{Nd})$ with identical one-body marginal probability densities $\mu_A$,
\be \label{gammaAmarg}
  \int_{\R^{d(N-1)}} \gamma_A(x_1,..,x_N) \, d\xhat_i = \mu_A(x_i) \;\;\; (i=1,..,N).
\ee
Note that in this subsection it is convenient to work with the above marginal probability density $\mu_A$, which integrates to $1$, rather than the one-body density $\rho_A=N\mu_A$, which integrates to the number $N$ of particles. 
Moreover we suppose that we are given a ``target marginal'', i.e. a second probability density $\mu_B\in\calP(\R^d)\cap L^1(\R^d)$. The goal is to construct a ``projection operator'' $P$ which maps $\gamma_A$ to a transportation plan $\gamma_B\in \calP(\R^{dN})\cap L^1(\R^{dN})$ with equal marginals $\mu_B$. As in \cite{CFK11}, we begin by defining the following transport plan on $\R^{2d}$ 
\be \label{gammaBA}
  \begin{array}{c}
   \gamma_{B,A}(x,x') := f(x)\delta_x(x') + \frac{f_B(x)f_A(x')}{\int f_B}, \mbox{ where} \\
   f(x):=\min\{\mu_A(x),\mu_B(x)\}, \, f_A:=(\mu_A-f)_+, \, 
   f_B :=(\mu_B-f)_+.
  \end{array} 
\ee
Here $h_+$ denotes the positive part of a function $h$, i.e. $h_+(x)=\max\{h(x),0\}$.
Clearly, $\mu_A=f+f_A$, $\mu_B=f+f_B$, $\int f_A = \int f_B = \frac12 ||\mu_A-\mu_B||_{L^1}$, and
\be \label{gammaBAmarg}
  \int \gamma_{B,A}(x,x')dx' = \mu_B(x), \;\; \int \gamma_{B,A}(x,x')dx = \mu_A(x).
\ee
We now define 
\be \label{P}
  P\gamma_A(x_1,..,x_N) := \int_{\R^d} \cdots \int_{\R^d} 
  \frac{\gamma_{B,A}(x_1,x_1' )}{\mu_A(x_1')} \cdots 
  \frac{\gamma_{B,A}(x_N,x_N' )}{\mu_A(x_N')} \; \gamma_A(x_1',..,x_N') \; dx_1' ...dx_N'. 
\ee
The operator $P$ of course depends on $\mu_A$ and $\mu_B$, which is suppressed in the notation.
By \eqref{gammaBAmarg} and the fact that $\gamma_A$ has equal marginals $\mu_A$, 
\be \label{PgammaAmarg}
   \int_{\R^{d(N-1)}} P\gamma_A \, d\xhat_i = \mu_B(x_i);
\ee
in particular $P\gamma_A\in \calP(\R^{Nd})\cap L^1(\R^{Nd})$. The following basic estimate serves to justify the above construction, although we will bypass it via a more elementary argument in the asymptotic analysis of the Coulomb cost.
\begin{proposition} \label{P:L1est} ($L^1$ stability) For any $\gamma_A\in\calP(\R^{Nd})\cap L^1(\R^{Nd})$ with identical one-body marginals $\mu_A$ (see \eqref{gammaAmarg}), and any $\mu_B\in\calP(\R^d)\cap L^1(\R^d)$, we have
$$
  ||\gamma_A - P\gamma_A||_{L^1(\R^{Nd})} \le c_N ||\mu_A - \mu_B||_{L^1(\R^d)},
$$
for some constant $c_N$ depending only on the number $N$ of particles.
\end{proposition}
{\bf Proof} Substituting the definition \eqref{gammaBA} of $\gamma_{B,A}$ into \eqref{P} and expanding yields the representation 
\begin{eqnarray*}
 P\gamma_A(x_1,..,x_N) &=& \prod_{i=1}^N \frac{f(x_i)}{\mu_A(x_i)} \gamma_A(x_1,..,x_N) \\
     &+& \sum_{i=1}^N \frac{f_B(x_i)}{\int f_B} \int \frac{f_A(x_i')}{\mu_A(x_i')} \gamma_A(x_i', \xhat_i) dx_i' \prod_{j\in\{1,..,N\}\backslash\{ i\}} \frac{f(x_j)}{\mu_A(x_j)} \\
     &+& \sum_{i,j=1}^N \frac{ f_B(x_i) f_B(x_j)}{(\int f_B)^2} 
         \int\int \frac{f_A(x_i')f_A(x_j')}{\mu_A(x_i')\mu_A(x_j')} \gamma_A(x_i',x_j',\xhat_{ij})\, dx_i' dx_j' \prod_{k\in\{ 1,..,N\}\backslash\{ i,j\} } \frac{f(x_k)}{\mu_A(x_k)} \\
     &+& ... \;\; + \;\; \prod_{i=1}^N \frac{f_B(x_i)}{\int f_B} \int\cdots\int 
         \prod_{k=1}^N \frac{f_A(x_k')}{\mu_A(x_k')} \gamma_A(x_1',..,x_N')dx_1'...dx_N',
\end{eqnarray*}
where $\xhat_{ij}$ denotes the coordinates $(x_k)_{k\neq i,j}$. This expansion can be written more compactly as a sum over multi-indices $I\subseteq\{1,..,N\}$: letting $x_I'=(x_i')_{i\in I}$, $I^c=\{1,..,N\}\backslash I$, $x_{I^c}=(x_k)_{k\in I^c}$, we have
\be \label{Pexpan}
  P\gamma_A(x_1,..,x_N) = \!\!\! \sum_{I\subseteq\{1,..,N\}} \!\!\! P_I\gamma_A, \;\;
  P_I\gamma_A = \prod_{i\in I} \frac{f_B(x_i)}{\int f_B} \int\! \cdots \!\int 
  \prod_{i\in I} \frac{f_A(x_i')}{\mu_A(x_i')} \gamma_A(x_I',x_{I^c}) \, dx_I'
  \prod_{k\in I^c} \frac{f(x_k)}{\mu_A(x_k)}.  
\ee
For $|I|\ge 1$, we estimate $P_I\gamma_A$ as follows. Estimating $f/\mu_A\le 1$, 
and $f_A(x_i')/\mu_A(x_i')\le 1$ for all but one $i\in I$, say $i\in I\backslash\{i_0\}$, integrating over $x_{I^c}$, and splitting the integration over $x_I'$ into first integrating over $x_{I\backslash\{i_0\}}'$ then over $x_{i_0}'$ gives
$$
 \int P_I \gamma_A(x) dx_{I^c} \le \prod_{i\in I} \frac{f_B(x_i)}{\int f_B} \;
     \int \frac{f_A(x_{i_0}')}{\mu_A(x_{i_0}')} 
     \Bigl[ \int \gamma_A(x_I',x_{I^c})\, dx'_{I\backslash\{i_0\}} dx_{I^c} \Bigr] dx_{i_0}'.
$$
But the integral in the square brackets equals $\mu_A(x_{i_0}')$, whence 
$$
  \int P_I\gamma_A(x) dx_{I^c} \le \prod_{i\in I} \frac{ f_B(x_i)}{\int f_B} \int f_A(x_{i_0}') dx_{i_0}'.
$$
Integrating over $x_I$ yields 
\be \label{PI}
   \int P_I\gamma_A(x)dx \le \int f_A.
\ee
For $I=\emptyset$, i.e. the first term in \eqref{Pexpan}, we have
\begin{eqnarray*}
   (\gamma_A - P_\emptyset \gamma_A)(x_1,..,x_N) &=& \Bigl( 1 - \prod_{i=1}^N \frac{f(x_i)}{\mu_A(x_i)} \Bigr)
   \gamma_A \\
   &=& \Bigl(1 - \frac{f(x_i)}{\mu_A(x_i)} \Bigr) \prod_{i=2}^N \frac{f(x_i)}{\mu_A(x_i)} \gamma_A 
   +\Bigl(1 - \prod_{i=2}^N\frac{f(x_i)}{\mu_A(x_i)} \Bigr) \gamma_A
\end{eqnarray*}
and hence, by iteration, 
$$
  \gamma_A - P_\emptyset\gamma_A = \sum_{k=1}^N
  \Bigl(1 - \frac{f(x_k)}{\mu_A(x_k)}\Bigr) \prod_{i>k} \frac{f(x_i)}{\mu_A(x_i)} \, \gamma_A
$$
(with the convention that the product $\prod_{i>k}$ equals $1$ when $k=N$).
Estimating the factors $f(x_i)/\mu_A(x_i)$ by $1$ and using, for the $k^{th}$ term, that 
$\int \gamma_A d\xhat_k = \mu_A(x_k)$ gives
\be \label{Pempty}
 \int(\gamma_A - P_\emptyset\gamma_A)dx \le \sum_{k=1}^N \int(\mu_A(x_k)-f(x_k))dx_k = N\int f_A.
\ee
Combining \eqref{PI}, \eqref{Pempty} and using that there are $2^N-1$ $I$'s with $|I|\ge 1$ yields
$$
   ||\gamma_A - P\gamma_A||_{L^1} \le (2^N-1+N)\int f_A = (2^{N-1}+\frac{N-1}{2})||\mu_A-\mu_B||_{L^1},
$$
establishing the proposition. 
\\[2mm]
We now state the main technical result of this section.
\begin{theorem}\label{T:Sobpres} (Preservation of Sobolev regularity) Let $\gamma_A\in\calP(\R^{Nd})\cap L^1(\R^{Nd})$, with equal one-body marginals $\mu_A$, and let 
$\mu_B\in\calP(\R^d)\cap L^1(\R^d)$. If $\sqrt{\gamma_A}\in H^1(\R^{Nd})$, $\sqrt{\mu_B}\in H^1(\R^d)$, and $\gamma_A$ is strongly positive (see Definition \eqref{D1}), then the plan $P\gamma_A$ defined by \eqref{P} satisfies $\sqrt{P\gamma_A}\in H^1(\R^{Nd})$. 
\end{theorem}
If, in addition, $\gamma_A$ is symmetric, 
this result, together with eq. \eqref{PgammaAmarg} and the fact that $P\gamma_A$ remains symmetric, shows that $P\gamma_A$ is the $N$-body density \eqref{Nbodydensbos} of a bosonic wavefunction $\Psi\in\calb_N$ with one-body density $\rho_B=N\mu_B$, namely $\Psi = \sqrt{P\gamma_A}$.

We also note that if any transportation plan has a square root belonging to $H^1$, then so does
its marginal. This is a consequence of the following well-known elementary estimate going back to \cite{HO77}:
\begin{lemma} \label{L:A4} (Hoffmann-Ostenhof inequality) If $u\in H^1(\R^n)$ and $\mu_k(x)=\int_{\R^{n-k}}|u(x,y)|^2 dy$, then
$$
     \int_{\R^k} |\nabla\sqrt{\mu_k}|^2 \le \int_{\R^n} |\nabla_x u(x,y)|^2 dx \, dy.
$$ 
\end{lemma}
(To see this, note $\nabla\sqrt{\mu_k}(x) = (1/\sqrt{\mu_k(x)}) \, \mbox{Re} \int_{\R^{n-k}}\overline{u(x,y)}\nabla_x u(x,y)\, dy$, estimate the integral by the Cauchy-Schwarz inequality, square both sides, and integrate over $x$.)

Hence we conclude, first of all, that the assumption on $\gamma_A$ in Theorem \ref{T:Sobpres} implies $\sqrt{\mu_A}\in H^1$, and second, that the assumption $\sqrt{\mu_B}\in H^1$ is in fact necessary. 
\\[2mm]
{\bf Proof} It is clear that $\sqrt{P\gamma_A}\in L^2(\R^{Nd})$, since we have already shown that $P\gamma_A\in L^1(\R^{Nd})$. We need to show that $\nabla_{x_i}\sqrt{P\gamma_A}\in L^2(\R^{Nd})$ for all $i$, and may without loss of generality assume $i=1$. It is useful to write $P\gamma_A$ in a somewhat different form as compared to \eqref{Pexpan}: using the explicit form \eqref{gammaBA} of $\gamma_{B,A}(x_1,x_1')$ in \eqref{P} yields the following decomposition into two terms, 
\begin{eqnarray}
  P\gamma_A(x_1,..,x_N) &=& \frac{f(x_1)}{\mu_A(x_1)} 
            \underbrace{\int \prod_{i=2}^N \frac{\gamma_{B,A}(x_1,x_1' )}{\mu_A(x_1')}
            \gamma_A(x_1,x_2',..,x_N') }_{=:g(x_1,..,x_N)} \nonumber \\
  &+& \frac{f_B(x_1)}{\int f_B} \underbrace{\int \frac{f_A(x_1')}{\mu_A(x_1')} 
             \prod_{i=2}^N \frac{\gamma_{B,A}(x_i,x_i' )}{\mu_A(x_i')}
             \gamma_A(x_1',..,x_N')\, dx_1'...dx_N'.}_{=:\tilde{g}(x_2,..,x_N)} 
  \label{Pexpan2}
\end{eqnarray}
Differentiating with respect to $x_1$ gives 
\begin{eqnarray}
 \nabla_{x_1}\sqrt{P\gamma_A} &=& \frac{1}{2\sqrt{P\gamma_A}} \Bigl(
      \frac{\nabla f}{\mu_A}\, g - \frac{f \, \nabla\mu_A}{\mu_A^2} g 
      + \frac{f}{\mu_A} \nabla_{x_1}g + \frac{\nabla f_B}{\int f_B} \, \tilde{g}\Bigr)\nonumber \\
      &=:& W_1  + W_2 + W_3 + W_4. \label{grad} 
\end{eqnarray}
We will show that each of these terms belong to $L^2$. 
We will use the following lower bounds, the first of which is obvious from \eqref{Pexpan2} but the second of which is more subtle:
\be \label{5.22}
   P\gamma_A(x_1,..,x_N) \ge \frac{f(x_1)}{\mu_A(x_1)} g(x_1,..,x_N)
\ee
and
\be \label{5.28}
   P\gamma_A(x_1,..,x_N) \ge \beta \mu_B(x_1)\tilde{g}(x_2,..,x_N).
\ee
To prove \eqref{5.28}, note that by the strong positivity of $\gamma_A$,
$$
    \gamma_A(x_1',..,x_N') \ge \beta \mu_A(x_1')  
    \underbrace{\int \gamma_A(x_1',..,x_N')dx_1'}_{=:M_{2,..,N}\gamma_A(x_2',..,x_N')}.
$$
Multiplying by $\prod_{i=1}^N(\gamma_{B,A}(x_i,x_i')/\mu_A(x_i'))$ and integrating over the $x_i'$ yields
\be \label{Pbd1}
   P\gamma_A(x_1,..,x_N) \ge \beta (P\mu_A)(x_1)(PM_{\hat{1}}\gamma_A)(x_2,..,x_N)
   = \beta \mu_B(x_1) (PM_{2,..,N}\gamma_A)(x_2,..,x_N).
\ee
On the other hand, from the definition of $\tilde{g}$ we obtain, by estimating 
$f_A/\mu_A\le 1$ and using 
$\int\gamma_A(x_1',..,x_N')dx_1'=M_{2,..,N}\gamma_A(x_2',..,x_N')$,
\be \label{Pbd2}
  \tilde{g}(x_2,..,x_N) \le PM_{2,..,N}\gamma_A(x_2,..,x_N).
\ee
Combining \eqref{Pbd1}, \eqref{Pbd2} yields \eqref{5.28}. Finally, we will use that due to 
$\int \gamma_{B,A}(x_i,x_i')dx_i=\mu_A(x_i')$,
\be \label{gid}
   \int g(x_1,..,x_N)dx_2...dx_N = \int \gamma_A(x_1,x_2',..,x_N')dx_2'...dx_N' = \mu_A(x_1)
\ee
and
\be \label{gtildeid}
  \int \tilde{g}(x_2,..,x_N)dx_2...dx_N = \int \frac{f_A(x_1')}{\mu_A(x_1')}
     \Bigl( \int \gamma_A(x_1',..,x_N')dx_2'...dx_N'\Bigr) dx_1' = \int f_A.
\ee
We can now estimate the four terms in \eqref{grad}. For $W_1$, we use the lower bound \eqref{5.22}, giving
$$
 |W_1| \le \frac12 \sqrt{\frac{\mu_A}{f g}} \, \frac{|\nabla f|}{\mu_A} \, g = (\nabla\sqrt{f}) \, \sqrt{\frac{g}{\mu_A}}.
$$
Squaring, integrating over $x_2,..,x_N$, and using \eqref{gid} yields
$$
  \int |W_1(x_1,..,x_N)|^2 dx_2...dx_N \le |\nabla\sqrt{f}(x_1)|^2 \frac{\int g(x_1,
  ..,x_N)dx_2...dx_N}{\mu_A(x_1)} = |\nabla\sqrt{f}(x_1)|^2.
$$
Since $\nabla\sqrt{f} = \chi_{\mu_A>\mu_B}\nabla\sqrt{\mu_B} + \chi_{\mu_A\le \mu_B}\nabla\sqrt{\mu_A}$, $\sqrt{f}\in H^1(\R^d)$, so the right hand side is in $L^1(\R^d)$, showing that $W_1\in L^2(\R^{Nd})$. 

For $W_2$, using first the lower bound \eqref{5.22} and then the bound $\sqrt{f/\mu_A}\le 1$ gives
$$
   |W_2| \le \frac12 \sqrt{\frac{\mu_A}{fg}} \, |\nabla\mu_A| \, \frac{fg}{\mu_A^2} = |\nabla\sqrt{\mu_A}| \, \frac{\sqrt{fg}}{\mu_A} \le |\nabla\sqrt{\mu_A}| \, \sqrt{\frac{g}{\mu_A}}.
$$
Squaring and integrating over $x_2,..,x_N$ gives, thanks to \eqref{gid},
$$
  \int |W_2(x_1,..,x_N)|^2 dx_2...dx_N \le |\nabla\sqrt{\mu_A}(x_1)|^2. 
$$
Again, the right hand side belongs to $L^1(\R^d)$, establishing that $W_2\in L^2(\R^{Nd})$. 

Next, we deal with $W_3$. We write $\nabla_{x_1}g$ as follows, abbreviating $d\xhat_1'=dx_2'...dx_N'$
$$
 \nabla_{x_1}g = \int 2(\nabla_{x_1}\sqrt{\gamma_A(x_1,\xhat_1')}) \,  \sqrt{\gamma_A(x_1,\xhat_1')} \, \prod_{i=2}^N \frac{\gamma_{B,A}(x_i,x_i')}{\mu_A(x_i')} d\xhat_1'.
$$
The Cauchy-Schwarz inequality yields 
$$
  |\nabla_{x_1}g|^2 \le 4 \int |\nabla_{x_1}\sqrt{\gamma_A}(x_1,\xhat_1')|^2 \prod_{i=2}^N 
  \frac{\gamma_{B,A}(x_i,x_i')}{\mu_A(x_i')} d\xhat_1' \;
  \underbrace{\int \gamma_A(x_1,\xhat_1')\prod_{i=2}^N 
  \frac{\gamma_{B,A}(x_i,x_i')}{\mu_A(x_i')} d\xhat_1'}_{=g(x_1,x_2,..,x_N)}.
$$
Combining this with the lower bound \eqref{5.22} and cancelling factors which appear in both numerator and denominator gives 
$$
  |W_3|^2 \le \frac{f}{\mu_A} 
  \int |\nabla_{x_1}\sqrt{\gamma_A}(x_1,\xhat_1')|^2 \prod_{i=2}^N 
  \frac{\gamma_{B,A}(x_i,x_i')}{\mu_A(x_i')} d\xhat_1'.
$$
Estimating $f/\mu_A\le 1$, integrating over $x_2,..,x_N$ and using $\int\gamma_{B,A}(x_i,x_i')dx_i = \mu_A(x_i')$ yields
$$
  \int |W_3(x_1,..,x_N)|^2 dx_2...dx_N \le \int |\nabla_{x_1}\sqrt{\gamma_A}(x_1,\xhat_1')|^2 d\xhat_1'. 
$$
Finally, since $\sqrt{\gamma_A}\in H^1(\R^{Nd})$, the right hand side belongs to $L^1(\R^d)$. This shows that $W_3\in L^2(\R^{Nd})$. 

It remains to deal with $W_4$. Estimating the term $1/\sqrt{P\gamma_A}$ via the naive lower bound on $P\gamma_A$ obtained by keeping only the second term in \eqref{Pexpan2} does not work, as was explained in \cite{CFK11}, Section 5.3. Instead we need the concept of strong positivity introduced in Def. \ref{D1}, and the ensuing more subtle lower bound \eqref{5.28}. Consequently
\begin{eqnarray}
 |W_4| & \le & \frac{1}{2\sqrt{\beta \mu_B(x_1)\tilde{g}(x_2,..,x_N)}} \, \frac{|\nabla f_B(x_1)|}{\int f_B} \, \tilde{g}(x_2,..,x_N) \nonumber \\
 & = & \frac{1}{\sqrt{\beta}\int\! f_B} \, \frac{|\nabla f_B(x_1)|}{2\sqrt{\mu_B(x_1)}} \, \sqrt{\tilde{g}(x_2,..,x_N)}. \label{W4est}
\end{eqnarray}
We now use that $f_B\neq 0$ only when $\mu_B>\mu_A$, and that consequently 
$\nabla f_B = \chi_{\mu_B>\mu_A}(\nabla\mu_B - \nabla \mu_A)$ a.e. It follows that
\be \label{newest}
  \frac{|\nabla f_B|}{2\sqrt{\mu_B}} \le \chi_{\mu_B>\mu_A}
  \frac{|\nabla\mu_A| + |\nabla\mu_B|}{2\sqrt{\mu_B}} \le 
  \frac{|\nabla\mu_A|}{2\sqrt{\mu_A}} + \frac{|\nabla\mu_B|}{2\sqrt{\mu_B}} 
  = |\nabla\sqrt{\mu_A}| + |\nabla\sqrt{\mu_B}|.
\ee
Substituting \eqref{newest} into\eqref{W4est}, squaring, and using the elementary bound
$(a+b)^2 \le 2(a^2+b^2)$ with $a=|\nabla\sqrt{\mu_A}|$, $b=|\nabla\sqrt{\mu_B}|$ gives 
$$
  |W_4|^2 \le \frac{2}{\beta(\int f_B)^2} \;
   \Bigl( |\nabla\sqrt{\mu_A}|^2 + |\nabla\sqrt{\mu_B}|^2 \Bigr) \; \tilde{g}(x_2,..,x_N).
$$
Integrating over $x_2,..,x_N$ and using \eqref{gtildeid} yields
$$
  \int |W_4(x_1,..,x_N)|^2 dx_2...dx_N \le \frac{2}{\beta\int f_B}
  \Bigl( |\nabla\sqrt{\mu_A}(x_1)|^2 + |\nabla\sqrt{\mu_B}(x_1)|^2 \Bigr). 
$$
Since $\sqrt{\mu_A}$, $\sqrt{\mu_B}$ belong to $H^1$, it follows that $W_4\in L^2(\R^{Nd})$, as required. This establishes that $\nabla_{x_1}\sqrt{P\gamma_A}\in L^2(\R^{Nd})$, and completes the proof of Theorem \ref{T:Sobpres}.
\subsection{Controlling the Coulomb cost}
Here we generalize our analysis of the Coulomb cost of the plan $P\gamma_A$ in \cite{CFK11}, Section 5.2 from two-body to $N$-body problems.
\begin{proposition} \label{P:Coul} Let $\gamma_A\in\calP_{sym}(\R^{3N})\cap L^1(\R^{3N})$ with equal marginals $\mu_A$, let $\mu_B\in\calP(\R^3)$, and assume that $\mu_A$, $\mu_B\in L^1\cap L^3(\R^3)$. Then
$$
  \tilde{V}_{ee}[P\gamma_A] \le \tilde{V}_{ee}[\gamma_A] + c_* {N\choose 2} ||\mu_A-\mu_B||_{L^1\cap L^3(\R^3)},
$$
where $||h||_{L^1\cap L^3}=\max\{ ||h||_{L^1}, \, ||h||_{L^3}\}$ and $c_*=6 (\frac{8\pi}{3})^{1/3}$.
\end{proposition}
{\bf Proof} From definition \eqref{P} it is elementary to check that $P$ commutes with taking $k$-body marginals, i.e.
\be \label{PCoulest}
   M_k P \gamma_A = P M_k \gamma_A,
\ee
where for any $k\in\{1,..,N-1\}$ and any $\gamma\in\calP_{sym}(\R^{3N})$, the $k$-body marginal $M_k\gamma$ is defined by 
\be \label{kmarg}
    (M_k\gamma)(A) = \gamma(A\times \R^{3(N-k)}) \mbox{ for all measurable }A\subseteq\R^{3k}.
\ee
The assertion now follows by using that for any $\gamma\in\calP_{sym}(\R^{3N})$
\be \label{2bodyred}
     \tilde{V}_{ee}[\gamma] = {N \choose 2} \tilde{V}_{ee}[M_2\gamma]
\ee
and applying the corresponding estimate for $N=2$ in \cite{CFK11}, Section 5.2.
\subsection{Smoothing} \label{sec:smoothing}
As in \cite{CFK11}, let $\phi_\eps(x)=\frac{1}{\eps^3}\phi(\frac{x}{\eps})$, where $\phi=\eta^2>0$ for some function $\eta$ belonging to the Schwartz space ${\cal S}(\R^3)$ of smooth, rapidly decaying functions, $\int\phi=1$, and $\phi$ radially symmetric. For instance, $\phi(x)=\pi^{-3/2}e^{-|x|^2}$ will do. For given $\gamma\in \calP_{sym}(\R^{3N})$ with $\gamma\mapsto\mu$, let
\be \label{smooth}
   \gamma_\eps(x_1,..,x_N) = (\phi_\eps\otimes \cdots \otimes\phi_\eps *\gamma)(x_1,..,x_N)
   = \int_{\R^{3N}}\phi_\eps(x_1-x_1')\cdot ... \cdot \phi_\eps(x_N-x_N')\, d\gamma(x_1',..,x_N').
\ee
\begin{lemma} \label{L:smooth} For any $\gamma\in \calP_{sym}(\R^{3N})$, the mollified density $\gamma_\eps$ introduced above satisfies \\
(i) $\gamma_\eps\in C^{\infty}(\R^{3N})$, $\gamma_\eps$ symmetric, $\gamma_\eps>0$, $\int\gamma_\eps=1$. \\
(ii) For any $k=1,..,N-1$, the $k$-point marginal $M_k\gamma_\eps$ of $\gamma_\eps$ is the mollification (\eqref{smooth} with $N$ replaced by $k$) of the $k$-point marginal $M_k\gamma$ of $\gamma$, i.e. $M_k \gamma_\eps = (M_k\gamma)_\eps$. In particular, the one-point marginal of $\gamma_\eps$ is 
$$
     \mu_\eps(x) = (\phi_\eps *\mu)(x) = \int_{\R^3} \phi_\eps(x-x')\, d\mu(x').
$$
(iii) $\tilde{V}_{ee}[\gamma_\eps]\le \tilde{V}_{ee}[\gamma]$ \\
(iv) $\sqrt{\gamma_\eps}\in H^1(\R^{3N})$.
\end{lemma}
The proof of (i), (ii), (iv) is a straightforward adaptation of the proof of Proposition 5.11 in \cite{CFK11}, and hence omitted. The Coulombic inequality (iii), which relies on Newton's theorem of electrostatics, follows by {\it applying} the corresponding result for $N=2$ in \cite{CFK11} to the two-body marginal $M_2\gamma$. Indeed, by \eqref{2bodyred}, (ii), and the $N=2$ result,
$$
    \tilde{V}_{ee}[\gamma_\eps] = {N \choose 2} \tilde{V}_{ee}[(M_2\gamma)_\eps] \le 
    {N\choose 2} \tilde{V}_{ee}[M_2\gamma] = \tilde{V}_{ee}[\gamma],
$$
establishing (iii).
\subsection{Passage to the limit} \label{sec:passage}
We are now in a position to prove Theorem \ref{T:bosons}. The arguments are analogous to those given for $N=2$ in \cite{CFK11}, Section 5.5. For completeness we include the details. 
\\[2mm]
{\bf Proof} We start from any fixed $\gamma\in\calP_{sym}(\R^{3N})$ with $\gamma\mapsto\mu=\rho/n$, where $\sqrt{\rho}\in H^1(\R^3)$. Let $\gamma_\eps$ be its mollification \eqref{smooth}. For $\beta\in(0,1)$, let $\gamma_{\eps,\beta}$ be the following ``strong positivization'' of $\gamma_\eps$, obtained by mixing in a small amount of a ``partial mean field plan'':  
\be \label{epsbeta}
   \gamma_{\eps,\beta}(x_1,..,x_N) = (1-\beta)\gamma_\eps(x_1,..,x_N) + \frac{\beta}{N}
   \sum_{i=1}^N M_1\gamma_\eps(x_i) M_{N-1}\gamma_\eps(\xhat_i).
\ee
Here $M_1\gamma_\eps$ and $M_{N-1}\gamma_\eps$ are the $1$-body and $(N-1)$-body marginals introduced in \eqref{kmarg}. Obviously $\gamma_{\eps,\beta}$ is strongly positive (with constant $\beta'=(1-\beta)\beta/N$), and has the same one-body marginal $\mu_\eps=M_1\gamma_\eps$ as $\gamma_\eps$. We also note that 
\be \label{M2}
 M_2\gamma_{\eps,\beta} = (1-\beta)M_2\gamma_\eps + \frac{\beta}{N}M_2(\sum_{i=1}^N M_1\gamma_\eps(x_i)M_{N-1}\gamma_\eps(\xhat_i))
 = \Bigl( 1-\frac{2\beta}{N}\Bigr) M_2(\gamma_\eps) + \frac{2\beta}{N}\mu_\eps\otimes\mu_\eps,
\ee
the last equality being due to the fact that the two-point marginal of the terms in the sum equals $\mu_\eps\otimes\mu_\eps$ when $i=1,2$ and $M_2\gamma_\eps$ for $i=3,..,N$. 

By Theorem \eqref{T:Sobpres} with $\mu_A=\mu_\eps$ and $\mu_B=\mu$, the plan $P\gamma_{\eps,\beta}$ has one-body marginal $\mu$ and satisfies $\sqrt{P\gamma_{\eps,\beta}}\in H^1(\R^{3N})$, and is hence an admissible trial function in the variational principle \eqref{FBos}. Thus we have
\be \label{Veeest0}
   F_\alpha^{Bos}[\rho] \le \alpha T[\sqrt{P\gamma_{\eps,\beta}}] 
                         + \tilde{V}_{ee}[P\gamma_{\eps,\beta}]
\ee
and hence, by finiteness of the kinetic energy $T[\sqrt{P\gamma_{\eps,\beta}}]$,
\be \label{Veeest1}
   \lim_{\alpha\to 0} \alpha T [P\gamma_{\eps,\beta}]=0.
\ee
By Proposition \eqref{P:Coul}, \eqref{M2}, \eqref{2bodyred}, and Lemma \ref{L:smooth} (iii) applied to both $\gamma_\eps$ and $(\mu\otimes\mu)_\eps=\mu_\eps\otimes\mu_\eps$, 
\begin{eqnarray} 
   \tilde{V}_{ee}[P\gamma_{\eps,\beta}] 
   & \le & \tilde{V}_{ee}[\gamma_{\eps,\beta}] + c_* {N\choose 2} ||\mu_\eps-\mu||_{L^1\cap L^3(\R^3)} \nonumber \\ 
   &  = & \Bigl( 1-\frac{2\beta}{N}\Bigr) \tilde{V}_{ee}[\gamma_\eps] + \beta(N-1)\tilde{V}_{ee}[\mu_\eps\otimes\mu_\eps] + c_* {N\choose 2} ||\mu_\eps-\mu||_{L^1\cap L^3(\R^3)}
   \nonumber \\
   & \le & \Bigl( 1-\frac{2\beta}{N}\Bigr) \tilde{V}_{ee}[\gamma] + \beta(N-1)\tilde{V}_{ee}[\mu\otimes\mu] + c_* {N\choose 2} ||\mu_\eps-\mu||_{L^1\cap L^3(\R^3)}. \label{Veeest2}
\end{eqnarray}
Since $\sqrt{\mu}\in H^1(\R^3)$, it belongs to $L^1\cap L^3(\R^3)$, so $\tilde{V}_{ee}[\mu\otimes\mu]$ is finite. Combining \eqref{Veeest0}, \eqref{Veeest1}, \eqref{Veeest2} and letting $\eps$ and $\beta$ tend to $0$ yields, thanks to the fact that the mollification $\mu_\eps$ of $\mu$ converges to $\mu$ in $L^1\cap L^3(\R^3)$,
\be \label{Veeest3}
   \limsup_{\alpha\to 0} F_\alpha^{Bos}[\rho] \le \tilde{V}_{ee}[\gamma].
\ee
Finally, since $\gamma$ was an arbitrary symmetric transportation plan with $\gamma\mapsto\mu=\rho/N$, taking the infimum over $\gamma$ gives
\be \label{Veeest4} 
  \limsup_{\alpha\to 0} F_\alpha^{Bos}[\rho] \le V_{ee}^{SCE}[\rho].
\ee
Since the reverse inequality $F_\alpha^{Bos}[\rho]\ge V_{ee}^{SCE}[\rho]$ is trivial, this establishes Theorem \ref{T:bosons}. 
\section{Proof of the semiclassical limit theorem, II: Fermions} \label{sec:fermions}
We now investigate the relationship between the bosonic and fermionic HK functionals \eqref{FBos}, \eqref{FHK}.

Our previous work \cite{CFK11} on the two-particle case relied on the fact that for $N=2$ the two functionals \eqref{FHK} and \eqref{FBos} coincide, since each symmetric wavefunction $\Psi\in\calb_N$ can be made antisymmetric by multiplication with an antisymmetric spin part (see the the proof of the proposition below). For $N\ge 3$
this is no longer the case:
\begin{proposition} \label{P:BosFer} 
a) For any $\rho\in\calR_N$, $F^{Bos}_\alpha[\rho] \le F^{HK}_\alpha[\rho]$. \\
b) (N=2) For any $\rho\in\calR_2$, $F^{Bos}_\alpha[\rho] = F^{HK}_\alpha[\rho]$. 
\end{proposition}
However, as we shall show via a careful analysis, the difference between
the two functionals disappears in the semiclassical
limit.  
\begin{theorem} \label{T:BosFer} For any $\rho\in\calR_N$ (see \eqref{densityclass}), 
$\lim_{\alpha\to 0} F^{HK}_\alpha[\rho] = \lim_{\alpha\to 0} F^{Bos}_\alpha[\rho]$.
\end{theorem}
{\bf Proof of the Proposition.} a) For any $\Psi\in\cala_N$ with $\Psi\mapsto\rho$, the function
$\Phi:=(\sum_{s_1,..,s_N}|\Psi(x_1,s_1,..,x_N,s_N)|^2)^{1/2}$ belongs to $\calb_N$ and satisfies $\Phi\mapsto\rho$, 
$V_{ee}[\Phi]=V_{ee}[\Psi]$. Moreover an argument similar to that in the proof of Lemma \ref{L:A4} shows
$T[\Phi]\le T[\Psi]$. \\[1mm]
b) Conversely, when $N=2$, for any $\Phi\in\calb_2$ with $\Phi\mapsto\rho$ the product of $\Phi$ with an antisymmetric spin function
belongs to $\cala_N$ and satisfies $\Psi\mapsto\rho$, $V_{ee}[\Psi]=V_{ee}[\Phi]$, $T[\Psi]=T[\Phi]$. 
\\[2mm]
Before coming to the proof of Theorem \ref{T:BosFer} we prepare two lemmas. 
\begin{lemma}\label{L:node} There exist two Lipschitz functions $A, \, B\, : \, \R^{3N}\to\R$ such that \\[1mm]
i) $A$ is antisymmetric, i.e. $A(x_{\sigma(1)},..,x_{\sigma(N)})=sgn(\sigma)A(x_1,..,x_N)$ for all permutations $\sigma$
\\[1mm]
ii) $A=1$ if $|x_i-x_j|>1$ for all $i\neq j$ 
\\[1mm]
iii) $A^2+B^2=1$.
\end{lemma}
{\bf Proof} Take two Lipschitz functions $c, \, s\, : \, \R\to\R$ such that $c^2+s^2=1$, $s$ odd, $s^2=1$
outside $[-1,1]$. For instance, 
\begin{eqnarray*}
    s(z) & := & \begin{cases} \sin(\frac{2}{\pi} z), & |z|<1 \\
                              \mbox{sign}(z),  & \mbox{otherwise} \end{cases} \\
    c(z) & := & \begin{cases} \cos(\frac{2}{\pi} z), & |z|<1 \\
                              0,  & \mbox{otherwise} \end{cases} 
\end{eqnarray*}
will do. Now take 
$$
   A(x_1,..,x_N) := \prod_{1\le i<j\le N} s\Bigl((x_i-x_j)_1)\Bigr),
$$
where $(x_i-x_j)_1$ denotes the first component of the vector $x_i-x_j\in\R^3$. It is easy to check
that $A$ is antisymmetric, and it is obvious that $A$ is Lipschitz. Define $B$ such that eq. (iii) holds, 
i.e. $B := \sqrt{1-A^2}$. It is not obvious that $B$ is Lipschitz, due to the appearance of the 
square root function, which is not Lipschitz. 

To investigate the matter, we analyze the gradient of $B$.
Let $M = {N\choose 2}$, let $s_1,..,s_M$ denote the functions $s((x_i-x_j)_1)$ ($1\le i<j\le N$), and 
let $c_1,..,c_M$ be the corresponding functions with $s$ replaced by $c$. Then $A=\prod_{\alpha=1}^Ms_\alpha$
and consequently 
$$
   B = \Bigl( \prod_{\alpha=1}^M(c_\alpha^2 + s_\alpha^2) - \prod_{\alpha=1}^M s_\alpha^2\Bigr)^{1/2}
     = \Bigl( \sum_{I\subseteq\{1,..,M\},\,|I|\ge 1} \prod_{\alpha\in I}c_\alpha^2 
               \prod_{\beta\in I^c} s_\beta^2\Bigr)^{1/2},
$$
where $I^c$ denotes the complement of $I$, i.e. $I^c = \{1,..,M\}\backslash I$. In the region where $B>0$, 
$$
   \nabla B = \frac{1}{B} \sum_{I\subseteq\{1,..,M\}, \, |I|\ge 1} 
       \Bigl( \sum_{\ell\in I} c_\ell \nabla c_\ell \prod_{\alpha\in I\backslash\{\ell\} } c_\alpha^2
               \prod_{\beta\in I^c} s_\beta^2 
             + \sum_{\ell\in I^c} s_\ell \nabla s_\ell \prod_{\alpha\in I} c_\alpha^2 \prod_{\beta\in I^c
               \backslash\{\ell\} } s_\beta^2 \Bigr) .
$$
For the term coresponding to the subset $I$, we estimate the denominator $B$ from below by 
$\prod_{\alpha\in I}c_\alpha \prod_{\beta\in I^c} |s\beta|$, and note that each of these factors
is cancelled by a corresponding factor in the numerator. This together with the fact that all remaining 
factors in the numerator other than the gradients are in absolute value bounded above by one implies that
$$
    |\nabla B| \le \sum_{I\subseteq\{1,..,M\}, \, |I|\ge 1} 
           \Bigl( \sum_{\ell\in I} |\nabla c_\ell| + \sum_{\ell\in I^c}|\nabla s_\ell|\Bigr). 
$$
Hence $\nabla B$ is bounded, establishing the assertion. The proof of the lemma is complete.
\begin{lemma} \label{L:rep} (Representing an arbitrary density by an antisymmetric wavefunction) 
Given any $\rho\in \calR_N$, there exists a real-valued antisymmetric $\Psi\in\cala_N$ with 
$\Psi\mapsto\rho$.
In particular, since $\Psi\in\cala_N$, $T[\Psi]<\infty$. 
\end{lemma}
This essentially follows from the results of Harriman \cite{Ha81} and Lieb \cite{Li83}: in the former
paper a real-valued antisymmetric $\Psi$ with $\Psi\mapsto\rho$ is constructed and in the latter paper it is 
shown that a similar complex-valued $\Psi$, also constructed by Harriman, has the required Sobolev regularity,
i.e. square-integrable gradient, provided $\sqrt{\rho}$ does. For completeness, and also because the Harriman
construction has a beautiful and hitherto unnoticed OT interpretation, a proof of Lemma \ref{L:rep} is 
included in an appendix. (See Theorem \ref{T:A2}, Example 2.)
\\[2mm]
{\bf Proof of Theorem \ref{T:BosFer}.} First we observe that
\be \label{limhbar}
   \lim_{\alpha\to 0} F^{Bos}_\alpha[\rho] = \inf_{\Phi\in\calb_N, \, \Phi\mapsto\rho} V_{ee}[\Phi], \;\;\;
   \lim_{\alpha\to 0} F^{HK}_\alpha[\rho] = \inf_{\Psi\in\cala_N, \, \Psi\mapsto\rho} V_{ee}[\Psi].
\ee
By (\ref{limhbar}) and Proposition \ref{P:BosFer} a), it suffices to show 
\be \label{RTP}
    \inf_{\Psi\in\cala_N, \, \Psi\mapsto\rho} V_{ee}[\Psi] \le 
    \inf_{\Phi\in\calb_N, \, \Phi\mapsto\rho} V_{ee}[\Phi]. 
\ee
We start from any fixed $\Phi\in\calb_N$ with $\Phi\mapsto\rho$. 
\\[2mm]
{\bf Step 1: Node insertion.} Let $A$, $B$ be the functions delivered by Lemma \ref{L:node}, let $\delta>0$,
and define $A_\delta$, $B_\delta \, : \, \R^{3N}\to\R$ by 
$$
    A_\delta(x) := A(\frac{x}{\delta}), \;\;\; B_\delta(x) := B(\frac{x}{\delta}).
$$
Define $\chi\, : \, \{\pm\frac12\}^N\to\C$ by 
$$
   \chi(s_1,..,s_N) := \prod_{i=1}^N \delta_{1/2}(s_i).
$$
Now set 
\be \label{Psidelta}
   \Psi_\delta(x_1,s_1,..,x_N,s_N) := A_\delta(x_1,..,x_N) \, \Phi(x_1,..,x_N) \, \chi(s_1,..,s_N).
\ee
Then $\Psi_\delta$ is antisymmetric and belongs to $H^1((\R^3\times\Z_2)^N)$. Moreover its $N$-body density
$\rho^N_{\Psi_\delta}$ (see \eqref{posden}) satisfies
\be \label{rhoNsqueeze}
    0 \le \rho^N_{\Psi_\delta} \le |\Phi|^2,
\ee
and, since $\rho_{\Psi_\delta}^N = |\Phi|^2$ if $|x_i-x_j|\ge \delta$ for all $i\neq j$, 
\be \label{rhoNconv}
    \rho^N_{\Psi_\delta} \to |\Phi|^2 \mbox{ a.e. }(\delta \to 0).
\ee
Since $\sum_{i<j}\frac{1}{|x_i-x_j|}|\Phi|^2\in L^1(\R^{3N})$, respectively $|\Phi(x_1,\cdot)|^2\in 
L^1(\R^{3N-3})$ for a.e. $x_1$, it follows by dominated convergence from \eqref{rhoNsqueeze}, \eqref{rhoNconv} that 
\be \label{rhoNconv2}
   \rho_{\Psi_\delta}^N \to |\Phi|^2 \mbox{ in } L^1(\R^{3N}), 
\ee
\be \label{Veeconv}
   V_{ee}[\Psi_\delta] = \int_{\R^{3N}} \sum_{i<j}\frac{1}{|x_i-x_j|} \rho^N_{\Psi_\delta} dx_1..dx_N
   \to \int_{\R^{3N}} \sum_{i<j}\frac{1}{|x_i-x_j|} |\Phi|^2 dx_1..dx_N = V_{ee}[\Phi] \;\; (\delta\to 0),
\ee
\be \label{rhoconv}
   \rho_{\Psi_\delta}(x_1) = \int_{\R^{3N-3}}\rho^N_{\Psi_\delta}(x_1,\cdot) \to
   \int_{R^{3N-3}} |\Phi(x_1,\cdot)|^2 = \rho(x_1) \mbox{ for a.e. }x_1 \;\; (\delta\to 0).
\ee
This does not yet establish \eqref{RTP}, since $\Psi_\delta\not\mapsto\rho$. 
\\[2mm]
{\bf Step 2: Regularity of excess density} Integrating (\ref{rhoNsqueeze}) over $x_2,..,x_N$ yields
$0\le \rho_{\Psi_\delta}\le \rho$, whence the excess density
\be \label{excess}
    \rho'_\delta := \rho - \rho_{\Psi_\delta}
\ee
is nonnegative. Even though both $\sqrt{\rho}$ and $\sqrt{\rho_{\Psi_\delta}}$ belong to
$H^1(\R^3)$ (the latter on account of Lemma \ref{L:A4}), it is not obvious whether their difference
$\rho'_\delta$ does. It is highly nonobvious whether or not $\sqrt{\rho'_\delta}\in H^1(\R^3)$. An elementary
example of two densities with square root in $H^1$ whose difference does not have a square root in $H^1$
was given in \cite{CFK11}, Ex. 5.2. 

This is why we prepared Lemma \ref{L:node}. Define 
$$
   \Phi'_\delta := B_\delta \Phi.
$$
It follows from $A_\delta^2 + B_\delta^2 = 1$ (see Lemma \ref{L:node} (iii)) that 
$$
   |\Phi'_\delta|^2 + \rho^N_{\Psi_\delta} = |\Phi|^2
$$
and hence, by taking marginals, that
$$
   \rho'_\delta = \rho_{\Phi'_\delta}.
$$
But by the Lipschitz continuity of $B_\delta$ implied by Lemma \ref{L:node}, 
$\Phi'_\delta\in H^1(\R^{3N})$, and so by Lemma \ref{L:A4}
\be \label{rho'reg}
    \sqrt{\rho'_\delta} = \sqrt{\rho_{\Phi'_\delta}} \in H^1(\R^3).
\ee
Finally we establish a strengthening of the convergence \eqref{rhoconv} which says that $\rho'_\delta\to 0$
a.e. as $\delta\to 0$. Since $0\le\rho'_\delta\le\rho$, and $\rho\in L^1\cap L^3(\R^3)$, 
dominated convergence implies
\be \label{rho'conv}
   \rho'_\delta \to 0 \mbox{ in }L^1\cap L^3(\R^3) \;\; (\delta\to 0).
\ee   
{\bf Step 3: Representability of excess density.} The excess wavefunction $\Phi'_\delta=B_\delta\Phi$
which represents $\rho'_\delta$ is not antisymmetric, so we replace it. By the regularity \eqref{rho'reg}, 
Lemma \ref{L:rep} is applicable and there exists a wavefunction $\Psi'_\delta \mapsto\rho'_\delta$ 
which is real-valued, antisymmetric, and belongs to $H^1((\R^3\times\Z_2)^N)$. Next, let us estimate
its electron-electron interaction energy. Because the wavefunction delivered by Lemma \ref{L:rep}
is a Slater determinant $|\varphi_1,..,\varphi_N\rangle$, introducing the one-body density matrix
$\gamma_{\Psi'_\delta}(x,s,x',s') := \sum_{i=1}^N\varphi(x,s)\overline{\varphi(x',s')}$ and utilizing
a standard identity from quantum chemistry (see e.g. \cite{SO96}) yields
\be \label{Veeest}
 V_{ee}[\Psi'_\delta] = \frac12 \int_{\R^6} \frac{\rho_{\Psi'_\delta}(x) \rho_{\Psi'_\delta}(x') - 
                        \sum_{s\in\Z_2}|\gamma_{\Psi'_\delta}(x,s,x',s)|^2 }{|x-x'|} dx \, dx' 
                     \le \frac12 \int_{\R^6} \frac{\rho_{\Psi'_\delta}(x) \rho_{\Psi'_\delta}(x')}{|x-x'|} 
                       dx \, dx'.
\ee
Next, we use the following estimate from \cite{CFK11}: if $f\in L^1(\R^3)$ and $g\in L^1\cap L^3(\R^3)$, then
\be \label{CFK1est}
   \Bigl| \int_{\R^6} \frac{1}{|x-y|}f(x)\, g(y) \, dx \, dy\Bigr| \le c_0 ||f||_{L^1} ||g||_{L^1\cap L^3},
\ee
with $c_0=2 (8\pi/3)^{1/3}$. Combining \eqref{Veeest}, \eqref{CFK1est} and \eqref{rho'conv} yields
\be \label{Vee'conv} 
    V_{ee}[\Psi'_\delta] \to 0 \;\; (\delta\to 0).
\ee
{\bf Step 4: Wavefunction matching.} In the previous steps,
we have decomposed $\rho$ into a main part $\rho_{\Psi_\delta}$, represented by an antisymmetric wavefunction
$\Psi_\delta$, and an excess part $\rho'_\delta$, represented by another antisymmetric wavefunction 
$\Psi'_\delta$. It is not obvious how to combine the two wavefunctions in order to represent the
total density $\rho=\rho_\delta + \rho'_\delta$, since the map $\Psi\mapsto\rho$ is quadratic. In particular,
taking $\Psi=\Psi_\delta + \Psi'_\delta$ will not work, due to occurence of an interference term. 

To proceed, it is crucial that both $\Psi_\delta$ (see \eqref{Psidelta}) and $\Psi'_\delta$ (see Step 3) 
are real-valued; consequently the complex-valued function 
$$
   \tilde{\Psi}_\delta := \Psi_\delta + i \Psi'_\delta
$$
satisfies 
\be \label{decomp}
      |\tilde{\Psi}_\delta|^2 = |\Psi_\delta|^2 + |\Psi'_\delta|^2.
\ee
The above identity yields $\rho_{\tilde{\Psi}_\delta} = \rho_{\Psi_\delta} + \rho_{\Psi'_\delta}$,
that is to say $\tilde{\Psi}_\delta \mapsto\rho$. Clearly also $\tilde{\Psi}_\delta\in\cala_N$, so
it is an admissible trial function in the left hand side of \eqref{RTP}. Finally, \eqref{decomp} yields
$V_{ee}[\tilde{\Psi}_\delta] = V_{ee}[\Psi_\delta] + V_{ee}[\Psi'_\delta]$, which together with 
\eqref{Veeconv}, \eqref{Vee'conv} implies
\be \label{Veeestfer}
   V_{ee}[\tilde{\Psi}_\delta] \to V_{ee}[\Phi] \;\; (\delta\to 0).
\ee
This establishes \eqref{RTP}, completing the proof of Theorem \ref{T:BosFer}. 
\section{Asymptotic behaviour of wavefunctions} \label{sec:WF}
It is a straightforward matter to deduce Theorem B from the results
of Sections \ref{sec:bosons} and \ref{sec:fermions}. \\[2mm]
{\bf Proof of Theorem B} Let $\{\Psi_\alpha\}_{\alpha>0}$ be any sequence of minimizers of \eqref{VP2}. The sequence of $N$-body densities $\rho_N^{\Psi_\alpha}$ (see \eqref{posden}) is then a sequence of probability densities with fixed one-body marginal $\rho/N$, and therefore tight. Hence after passing to a subsequence, $\rho_N^{\Psi_\alpha}$ converges weakly in $\calP(\R^{3N})$ to some $\gamma\in\calP(\R^3)$. By the antisymmetry of the $\Psi_\alpha$, their $N$-body densities are symmetric, hence so is $\gamma$. That the marginal condition is preserved in the limit follows from standard arguments in optimal transport theory (see e.g. the proof of Theorem 4.1 in \cite{Vi09}). That $\gamma$ is a minimizer of \eqref{VP1} now follows by combining Theorem A with the lower bound 
$\liminf_{\alpha\to 0} (\alpha T + V_{ee})[\Psi_\alpha] \ge \tilde{V}_{ee}[\gamma]$. The latter is immediate from the nonnegativity of $\alpha T$ and the well known fact (see, e.g., \cite{CFP15}) that $\tilde{V}_{ee}$ (and indeed any such functional with the Coulomb cost replaced by a nonnegative, lower semicontinuous function on $\R^{3N}$) is lower semicontinuous with respect to weak convergence of probability measures.
\\[2mm]
In fact our results in Sections \ref{sec:bosons} and \ref{sec:fermions} yield the following more general statement in the spirit of De Giorgi's Gamma convergence. The latter is the right notion of convergence of functionals associated with the asymptotic behaviour of minimizers; more precisely it implies convergence of minimizers provided the sequence of functionals is equi-coercive (see e.g. the survey article \cite{Br06}). Recall that the halfarrow $\rightharpoonup$ denotes weak convergence of probability measures.
\begin{theorem} \label{T:WF} (Gamma convergence) 
Let $\gamma\in\calP_{sym}(\R^{3N})$ with $\gamma\mapsto\rho/N$, where $\sqrt{\rho}\in H^1(\R^3)$. Then: \\
(i) (lower bound) Any sequence of wavefunctions $\Psi_\alpha\in\cala_N$ 
with $\Psi_\alpha\mapsto\rho$ and $\sum_{s_1,..,s_N}|\Psi_\alpha|^2 \rightharpoonup \gamma$ satisfies 
$$
      \liminf\limits_{\alpha\to 0} \Bigl(\alpha T +     V_{ee}\Bigr)[\Psi_\alpha] 
      \ge \tilde{V}_{ee}[\gamma].
$$
(ii) (recovery sequence) There exists a sequence of wavefunctions $\Psi_\alpha\in\cala_N$ with $\Psi_\alpha\mapsto\rho$ and $\sum_{s_1,..,s_N}|\Psi_\alpha|^2 \rightharpoonup \gamma$ which satisfies
$$
     \lim_{\alpha\to 0} \Bigl(\alpha T +  V_{ee}\Bigr)[\Psi_\alpha] = \tilde{V}_{ee}[\gamma].
$$
\end{theorem}
{\bf Proof} The lower bound was already established above, as our argument only used the assumptions of Theorem \ref{T:WF}. A sequence saturating this bound can easily be constructed using our estimates \eqref{Veeest1}, 
\eqref{Veeest2}, and \eqref{Veeestfer}, which were derived for an arbitrary plan $\gamma$ respectively an arbitrary bosonic wavefunction $\Phi$. More precisely:
Let $\gamma$ be any transportation plan satisfying the assumptions of Theorem \ref{T:WF}. Take any sequences $(\eps_j)_{j\in\N}$, $(\beta_j)_{j\in\N}$ in $(0,1)$ with $\eps_j\to 0$, $\beta_j\to 0$. For each $j$, let $\Phi_j=\sqrt{P\gamma_{\eps_j,\beta_j}}\in\calb_N$ be the bosonic wavefunction constructed in Section \ref{sec:bosons}, and for $\delta>0$ let $(\tilde{\Psi}_j)_{\delta} = (\Psi_j)_\delta + i (\Psi'_j)_\delta\in\cala_N$ be the associated antisymmetric wavefunction from Section \ref{sec:fermions} (see \eqref{Psidelta} and Step 3). Next, by \eqref{Veeestfer}, \eqref{rhoNconv2}, and \eqref{rho'conv}, we can choose $\delta=\delta_j\to 0$ such that 
\be \label{finalest}
    V_{ee}[(\tilde{\Psi})_{\delta_j}] \le V_{ee}[\Phi_j] + \frac{1}{j}, \;\; 
    || \sum_{s_1,..,s_N}|(\Psi_j)_{\delta_j}|^2 - |\Phi_j|^2 ||_{L^1} \le \frac{1}{j}, \;\;
    || \sum_{s_1,..,s_N}|(\Psi'_j)_{\delta_j}|^2 ||_{L^1} \le \frac{1}{j}.
\ee
Finally, we choose $\alpha_j>0$, $\alpha_j\to 0$ such that $\alpha_j T[(\tilde{\Psi}_j)_{\delta_j}] \le 1/j$. We claim that
\be \label{finalclaim}
   \limsup_{j\to\infty} (\alpha_j T + V_{ee})[(\tilde{\Psi}_j)_{\delta_j}] \le \tilde{V}_{ee}[\gamma], 
   \;\;\; \sum_{s_1,..,s_N}|(\tilde{\Psi}_j)_{\delta_j}|^2 \rightharpoonup \gamma,
\ee
as required. The first claim is clear from the first estimate in \eqref{finalest} and the fact that $\limsup_{j\to\infty} V_{ee}[\Phi_j] \le \tilde{V}_{ee}[\gamma]$, by \eqref{Veeest2} and the convergence of $\mu_\eps$ to $\mu$ in $L^1\cap L^3$. The second claim can be inferred as follows. We have $\gamma_{\eps_j,\beta_j}\rightharpoonup\gamma$ by standard properties of mollification, hence $P\gamma_{\eps_j,\beta_j}=|\Phi_j|^2\rightharpoonup\gamma$ by Proposition \ref{P:L1est} and the fact that $\mu_\eps\to \mu$ in $L^1$, and consequently 
$\sum_{s_1,..,s_N}|(\Psi_j)_{\delta_j}|^2\rightharpoonup \gamma$ by the second estimate in \eqref{finalest}. Moreover $\sum_{s_1,..,s_N}|(\Psi'_j)_{\delta_j}|^2\to 0$ in $L^1$ by the last estimate in \eqref{finalest}. This together with \eqref{decomp} establishes the second part of \eqref{finalclaim}, and completes the proof of Theorem \ref{T:WF}.

\section{A counterexample where minimization does not commute with taking the semiclassical limit} \label{sec:counter}
Lest the reader think that it is obvious that minimization of general functionals commutes with taking semiclassical limits, we present a counterexample.
The counterexample shares the following three features of the Levy-Lieb constrained-search problem \eqref{VP2}
\begin{itemize}
\item it is an integral of a quadratic expression in $\psi$ and $\nabla\psi$
\item minimization is subject to an infinite list of quadratic constraints
\item the minimizer of the functional with kinetic energy deleted has infinite kinetic energy.
\end{itemize}
The main mathematical difference to the Hohenberg-Kohn functional \eqref{FHK} is that 
\begin{itemize}
\item the constraint is a quadratic expression not just in $\psi$, but in $\psi$ and $\nabla \psi$.
\end{itemize}
The possibility of such a counterexample is closely related to the {\it Lawrentiev phenomenon} in the calculus of variations (see e.g. [Da08] Section 4.7 and the references therein). It says that mild additional regularity restrictions on the competing functions can push up the minimum of a functional not just infinitesimally, as would be suggested by approximation, but by a finite amount, called {\it Lawrentiev gap}. In case of our example, the mild additional regularity is related to (and even smaller
than) that occuring between the HK functional and its semiclassical limit, namely finite kinetic energy, i.e. square-integrable gradient, as opposed to just integrable
gradient.  

Our
example is built upon the well known fact going back to Mania [Ma34] that the functional $J$ emerging in (\ref{reformulation}) below exhibits a Lawrentiev phenomenon (see the discussion at the end of this Section). We do not claim that this example has any density-functional-theoretic meaning, but it illustrates that our result in Theorem A cannot follow from simple general considerations. 
\begin{example} \label{P:lawrentiev}
Let ${\mathcal A}$ denote the class of measurable functions $\psi = (\psi_1,..,\psi_7) \, : \, [0,1] \to \R^7$ whose first component
$\psi_1$ has integrable first derivative, and which satisfy the boundary conditions $\psi_1(0)=0$, $\psi_1(1)=1$. Consider the quadratic functional
$$
   I[\psi] + \eps^2 T[\psi] := \int_0^1 \psi_6(x)\, \psi_7(x) \, dx  \; + \; \eps^2 \int_0^1 \psi'_1(x)^2 dx
$$
subject to the quadratic constraints
$$
 \begin{pmatrix} \psi_2 \\ \psi_3 \\ \psi_4 \\ \psi_5 \\ \psi_6 \\ \psi_7 \end{pmatrix} = 
 \begin{pmatrix} \psi_1^2 \\ \psi_1\, \psi_2 \\ (\psi_1')^2 \\ \psi_1' \, \psi_4 \\ \psi_5^2 \\ (id-\psi_3)^2 \end{pmatrix},
$$
where $id$ denotes the identity map $id(x)=x$. We have 
\be \label{first}
   \lim_{\epsilon\to 0} \inf_{\psi\in {\mathcal A}} \Bigl( I[\psi] + \eps^2 T[\psi] \Bigr) \ge \frac12 \, (\frac78)^2 \, (\frac{3}{10})^5,
\ee
but 
\be \label{second}
   \inf_{\psi\in{\mathcal A}} I[\psi] =  0.
\ee
\end{example}
In other words, passage to the $\epsilon\to 0$ limit in the minimization problem (\ref{first}) cannot be achieved by just taking the naive (pointwise) limit of
the functional. 
\\[2mm]
To begin to understand this, note first that by setting $u:=\psi_1$
and expressing the remaining components of $\psi$ in terms of $u$ via the constraints, the functional becomes
\be \label{reformulation}
   I[\psi] + \eps^2 T[\psi] = \underbrace{\int_0^1 \Bigl(x-u(x)^3\Bigr)^2{u'(x)}^6 dx}_{=:J[u]} + \eps^2 \underbrace{\int_0^1 u'(x)^2 dx}_{=:T[u]},
\ee
and is to be minimized over the class ${\mathcal B}$ of scalar functions $u \, : \, [0,1]\to\R$ which are integrable with integrable first derivative (i.e.,
in technical language, which belong to the Sobolev space $W^{1,1}((0,1))$) and satisfy the boundary conditions 
\be \label{BC} \tag{BC}
   u(0)=0, \;\;\; u(1)=1. 
\ee 
The functional $J$, first introduced by Mania \cite{Ma34}, is quite innocent looking; the integrand is just a nonnegative polynomial in $x$, $u$, and $u'$. The assertions in Example \ref{P:lawrentiev} now reduce to the following lemma which is a minor modification of  results in [Ma34, Da08] (see the discussion at the end of this section). 
\begin{lemma} \label{L:lawrentiev}
The functionals $J$ and $T$ from (\ref{reformulation}) satisfy 
\be \label{first_u}
  \lim_{\epsilon\to 0} \inf_{u\in {\mathcal B}} \Bigl( J[u] + \eps^2 T[u] \Bigr) \ge \frac12 \, (\frac78)^2 \, (\frac{3}{10})^5,
\ee
but
\be \label{second_u}
   \inf_{u\in {\mathcal B}} J[u] =  0.
\ee
\end{lemma}
What is the intuition behind this counter-intuitive gap between (\ref{first_u}) and (\ref{second_u})? 

The functional $J[u]$ in (\ref{second_u}) can easily be minimized explicitly, by making
the integrand zero everywhere. The unique minimizer is $u_{opt}(x)=x^{1/3}$ and the first factor in the integrand, $(x-u^3)^2$, can be interpreted as a penalty for deviating
from this minimizer. 

On the other hand, this minimizer has infinite kinetic energy, because $u_{opt}'=\frac13 x^{-2/3}$, which is not square-integrable. This means that
the functional $J + \eps^2 T$ always has the value $+\infty$ on $u_{opt}$, no matter how small the constant $\eps>0$ is taken. This by itself does not explain the
gap yet, and is exactly analogous to the situation one has when computing the semiclassical limit of the Hohenberg-Kohn functional: the minimizer for $\epsilon=0$
has infinite kinetic energy. The question then is, is there a nearby function which has near-optimal $J$ but finite kinetic energy. In the HK case, our theorem shows that the answer is
Yes, but in the above example the answer is No. Why? Let us sketch the argument informally before making it rigorous. For small $x$, exact or approximate minimizers of the functional $J+\eps^2 T$ in (\ref{first_u}) can, due to the boundary condition $u(0)=0$ and finiteness of kinetic energy, at most grow like a constant times $x^{1/2}$ (see (\ref{csi})), and hence lie substantially below the faster-growing $\eps=0$ optimizer $u_{opt}=x^{1/3}$. But sooner or later they must start to grow away from zero, to reach the boundary value $u(1)=1$. Hence sooner or later the large penalty factor ${u'}^6$ in $J$ becomes active. A quantitative analysis, given below, then shows that the ensuing overall penalty $J[u]$ is bounded from below by the constant on the right hand side
of (\ref{first_u}). 

{\bf Proof of the lemma.} Let $\eps>0$ be arbitrary, and let $u$ be any function in ${\mathcal B}$ with $J[u]+\eps^2 T[u] < \infty$. In particular,
$T[u]<\infty$. In technical language, this means that $u$ belongs to the Sobolev space $W^{1,2}((0,1))$. It suffices to show that on such functions, $J$ is bounded away from zero, more precisely
\be \label{Jlowerboundfinal}
    \inf_{u\in {\mathcal B}\cap W^{1,2}((0,1))} J[u] \ge \frac12 \, (\frac78)^2 \, (\frac{3}{10})^5.
\ee
The idea is to estimate $J$ from below by a simpler functional whose minimizer
can be computed explicitly. By the boundary condition $u(0)=0$ and the Cauchy-Schwarz inequality,
\be \label{csi} 
   u(x) = \int_0^x u' \le \Bigl(\int_0^x 1\Bigr)^{1/2} \Bigl(\int_0^x {u'}^2\Bigr)^{1/2} \le x^{1/2} T[u]^{1/2}.
\ee
This implies that for small positive $x$, $u(x)<\frac12 u_{opt}(x)= \frac12 x^{1/3}$. On the other hand, for $x$ near $1$, by the boundary condition $u(1)=1$ we
have $u(x)>\frac12 u_{opt}(x)$. By the intermediate value theorem, there exists a smallest point $x_*\in(0,1)$ at which $u=\frac12 u_{opt}.$ (Note that functions $u\in {\mathcal B}$ are continuous.) Hence in $(0,x_*)$ 
\be
   x - u^3 = u_{opt}^3 - u^3 \ge u_{opt}^3 - (\frac12 u_{opt})^3 = \frac78 u_{opt}^3 = \frac78 x
\ee 
and consequently 
\be \label{Jlowerbound}
   J[u] \ge \int_0^{x_*} (x-u^3)^2 {u'}^6 \ge (\frac78)^2 \int_0^{x_*} x^2 u'(x)^6 \, dx =: G[u].
\ee
The lower bound functional $G$ no longer depends on $u$, and its minimizer subject to the boundary conditions $u(0)=0$, $u(x_*)=\frac12 u_{opt}(x_*)$ is easy
to find explicitly. The Euler-Lagrange equation is 
\be \label{eulerlagrangeG}
    x^2 u'^5 \equiv  C
\ee
for some constant $C$. Solving this equation subject to the first boundary condition shows that the minimizer is
given by $u_0(x)=\frac53 C^{1/5}x^{3/5}$. The constant $C$ can be determined from the second boundary condition, yielding
$$
    u_0(x) = \frac12 x^{3/5} x_*^{-4/15}.
$$
Evaluating $G$ explicitly on the above minimizer gives an explicit lower bound,
\be \label{Glowerbound}
   G[u] \ge G[u_0] = (\frac78)^2 \, (\frac3{10})^6 \, \frac53 x_*^{-1}.
\ee
Combining (\ref{Jlowerbound}), (\ref{Glowerbound}) and the fact that $x_*>1$ (since $x_*$ belongs to the interval $(0,1)$) yields (\ref{Jlowerboundfinal}), completing
the proof of Lemma \ref{L:lawrentiev} and the assertions in Example \ref{P:lawrentiev}. 
\\[2mm]
{\bf Discussion.} Equations (\ref{second_u}), (\ref{Jlowerboundfinal}) show that the functional $J$ exhibits a Lawrentiev gap, i.e. a gap of infima, between the function spaces $W^{1,1}((0,1))$ (integrable gradient) and $W^{1,2}((0,1))$ (square-integrable gradient). The analogous result with the second space replaced by the smoother space $W^{1,\infty}((0,1))$ (Lipschitz continuous functions) goes back to [Ma34] and is proved via arguments somewhat different from ours in [Da08]. 
Our reasoning shows that there remains such a gap between the function spaces $W^{1,p}((0,1))$ and $W^{1,q}((0,1))$ with $p<3/2$ and $q>3/2$. These spaces contain respectively
fail to contain the $\eps=0$ minimizer $u_{opt}=x^{1/3}$. 

The re-formulation as a quadratic problem with quadratic constraint given in Proposition \ref{P:lawrentiev} relies on the property of the Mania example $J$ that its integrand is a polynomial in $u$ and $u'$, 
and may be considered an application of the well known strategy in algebraic geometry to recast polynomial equations as systems of quadratic equations involving additional variables. 

The interesting question of what the minimizers of semiclassically perturbed functionals of type \eqref{reformulation} actually {\it do} in the presence of a Lawrentiev gap was first elucidated by Ball and Mizel \cite{BM84, BM84}, with the help of dynamical systems techniques. We note that the motivation of the latter authors to add the semiclassical term 
was very different from ours: making the Euler-Lagrange equation of functionals like 
$J$ elliptic yet showing that minimizers can still exhibit singularities, rather than understanding strong correlations in density functional theory. 
\appendix

\section{Appendix: Optimal transport interpretation of the Harriman-Lieb representation of a given density}
In 1981 Harriman \cite{Ha81} published several clever representations of a given density $\rho$ on $\R^3$ by orthonormal orbitals. Lieb \cite{Li83} analyzed the
regularity of the basic Harriman construction and proved that if the density satisfies $\sqrt{\rho}\in H^1(\R^3)$ -- as corresponding to the natural
domain of the Hohenberg-Kohn functional -- then the representing Slater determinant belongs to $H^1((\R^3\times\Z_2)^N)$ -- the natural domain of the
quantum energy functional. 

Since the Harriman representations fall out of thin air, it may be of interest to observe -- as we do in this Appendix -- that they are special cases
of a natural and general optimal transportation construction. As a straightforward offspring, we extend the Lieb regularity result to more general
representations, as was needed in Section \ref{sec:fermions}. 

Recall that any mapping $T \, : \, X \to Y$ between two sets $X$ and $Y$ which is onto naturally induces a linear mapping from functions on $X$ to functions on $Y$, defined by 
\be \label{pfdef}
   T_\sharp f(y) := f(T^{-1}(y)) \mbox{ for all }y\in Y.
\ee
The function $T_\sharp f$ is called the {\it push-forward} of $f$. 
\begin{lemma} \label{L:A1} Let $X\subseteq\R^m$, $Y\subseteq\R^n$ be Borel sets, and let $\mu$, $\nu$ be probability measures on $X$ respectively $Y$.
If $T \, : \, X\to Y$ is onto and pushes $\mu$ forward to $\nu$, i.e. $\mu(T^{-1}(A))=\nu(A)$ for all Borel $A\subseteq Y$, then $T_\sharp$ is a unitary mapping from $L^2(X,d\mu)$ to $L^2(Y,d\nu)$, that is to say 
$$
     \int_X f \, \overline{g} \, d\mu = \int_Y T_\sharp f \, \overline{T_\sharp g} \, d\nu
$$
for all $f, g\in L^2(X,d\mu)$. 
\end{lemma}
{\bf Proof} By the definition (\ref{pfdef}) of the push-forward of a function, the integral on the right equals $\int_Y f(T^{-1}(y)) \overline{g(T^{-1}(y))}\, d\nu(y)$. The result follows by changing variables and using that $T$ pushes $\mu$ forward to $\nu$. 
\\[2mm]
From now on we assume that $Y=\R^n$, and that $\nu$ is a given probability measure on $\R^n$ with density $v$ belonging to $L^1(\R^n)$. For $k=1,..,n$, we denote by $M_k\nu$ the marginal of $\nu$ with respect to the first $k$ variables, that is to say $M_k\nu(A) = \nu(A\times\R^{n-k})$ for all measurable $A\subseteq\R^3$. 
\begin{theorem} \label{T:A1} Let $X$ be a Borel subset of $\R^m$, and $\mu$ a probability measure on $X$. If $T \, : \, X \to\R^k$ is onto and pushes
$\mu$ forward to $M_k\nu$, then the map 
$$
    Lf(y) := \sqrt{v(y)} T_\sharp f(y_1,..,y_k)
$$
is a unitary map from $L^2(X,d\mu)$ to $L^2(\R^n, dy)$, where $dy$ denotes the Lebesgue measure on $\R^n$. 
\end{theorem}
Remark: Thus, to construct unitary embeddings from the general measure space $L^2(X,d\mu)$ into the standard Lebesgue space $L^2(\R^n)$,
it is not necessary to construct a map from $X$ to $\R^n$ which pushes $\mu$ forward to $\nu$, but it suffices to have a map from $X$ into a lower-dimensional subspace of $\R^n$ which pushes $\mu$ forward to the subspace marginal of $\nu$.
\\[2mm]
{\bf Proof} We write $y=(y_-,y_+)$ with $y_-=(y_1,..,y_k)$, $y_+=(y_{k+1},..,y_n)$. Substituting the definition of $L$, carrying out the integration 
over $y_+$, and applying Lemma \ref{L:A1} gives 
\begin{eqnarray*} 
    \int_{\R^n} Lf(y) \overline{Lg(y)} \, dy &=& \int_{\R^n} T_\sharp f(y_-) \overline{T_\sharp f(y_-)} \nu(y) \, dy \\
                                             &=& \int_{\R^k} T_\sharp f(y_-) \overline{T_\sharp f(y_-)}\, d(M_k\nu)(y_-) \\
                                             &=& \int_X f(x) \overline{g(x)} \, d\mu(x),
\end{eqnarray*}
as required. 
\begin{theorem} \label{T:A2} (Orthonormal orbitals in $\R^n$ with given density) Let $\nu$ be any given probability measure on $\R^n$ with density $v$ belonging to $L^1(\R^n)$. Let $X=[0,1]$, let $\mu=dx$ be the Lebesgue measure on $[0,1]$, and assume $T \, : \, [0,1]\to\R$ is onto
and pushes $dx$ forward to $M_1\nu$. Let $L \, : \, L^2([0,1]) \to L^2(\R^n)$ be the map from Theorem \ref{T:A1} with $k=1$, i.e. 
$$
     Lf(y) = \sqrt{v(y)} T_\sharp f(y_1).
$$
If $\varphi_1,..,\varphi_N$ \\
(i) are orthonormal functions in $L^2([0,1])$ \\
(ii) have constant denstiy, i.e. $\sum_{i=1}^N|\varphi_i(x)|^2 = N$, \\
then $L\varphi_1,..,L\varphi_N$ \\
(i') are orthonormal functions in $L^2(\R^{n})$ \\
(ii') have density $N \, v$, i.e. $\sum_{i=1}^N |L\varphi_i(y)|^2 = N\, v(y)$ for all $y\in\R^n$. 
\end{theorem}
{\bf Proof} Assertion (i') is immediate from Theorem \ref{T:A1}. To show (ii'), note first that, by substituting definitions, 
$$
    \sum_{i=1}^N |L\varphi_i(y)|^2 = v(y) T_\sharp \Bigl(\sum_{i=1}^N|\varphi_i|^2\Bigr)(y_1). 
$$
The assertion now follows from the fact that the push-forward of a constant function under any map is again a constant function, with the value of the function remaining unchanged. 
\\[2mm]
{\bf Example 1} (complex-valued Harriman-Lieb orbitals) Let $n=3$, $\rho\in L^1(\R^3)$, $\rho\ge 0$, $\int \rho = N$. 
Taking $\varphi_k(x)=e^{2\pi i k x}$, $\nu = v\, dy$ with $v=\rho/N$, and $T$ the standard map pushing $dx$ forward to $M_1\nu$, i.e. 
\be \label{standardmap}
    T = F^{-1} \mbox{ where }F(y_1) = \int_{-\infty}^{y_1} (M_1 v) (s)\, ds
\ee
where $M_1 v$ is the density of $M_1\nu$, i.e. $(M_1 v)(y_1)=\int_{\R^2} v(y_1,y_2,y_3) \, dy_2 dy_3$, gives 
$$
    L\varphi_k(y) = \sqrt{\frac{\rho(y)}{N}} e^{2\pi i k \, \int_{-\infty}^{y_1} \left(M_1 \frac{\rho}{N}\right)(s)\, ds }.
$$
Thease are the complex-valued 3D orbitals with density $\rho$ from \cite{Ha81} Section III respectively \cite{Li83} Section 1. 
\\[2mm]
{\bf Example 2} (real-valued Harriman orbitals) Let $n$, $\rho$, $\nu$, $T$ be as above, but take 
\begin{eqnarray*}
   & & \varphi_{2k-1}(x) = \sqrt{2}\sin 2\pi k x, \;\; \varphi_{2k}(x) = \sqrt{2}\cos 2 \pi k x \;\; (k\le \frac{N}{2}), \\
   & & \varphi_N(x) = 1 \mbox{ if $N$ is odd}. 
\end{eqnarray*}
Then 
\begin{eqnarray*}
   & & L\varphi_{2k-1}(y) = \sqrt{\frac{2\rho(y)}{N}}\sin \Bigl( 2\pi k \int_{-\infty}^{y_1} \Bigl(M_1 \frac{\rho}{N}\Bigr)(s)\, ds \Bigr),  
       L\varphi_{2k}(y) = \sqrt{\frac{2\rho(y)}{N}}\cos \Bigl( 2\pi k \int_{-\infty}^{y_1} \Bigl(M_1 \frac{\rho}{N}\Bigr)(s)\, ds \Bigr),  \\
   & & L\varphi_N(y) = \sqrt{\frac{\rho(y)}{N}} \mbox{ if $N$ is odd}. 
\end{eqnarray*}
These are the real-valued 3D orbitals with density $\rho$ from \cite{Ha81} Section III. 
\\[2mm]
Finally we investigate the regularity of the orbitals in Theorem \ref{T:A2}. The result below is a modest generalization of the result of Lieb [Lie83]
for the orbitals from Example 1.

\begin{theorem} \label{T:A3} (Regularity) Assume, in addition to the hypotheses of Theorem \ref{T:A2}, that the $\varphi_i$ are Lipschitz, and 
that $T$ is the standard map (\ref{standardmap}). If $\sqrt{v}$ belongs to $H^1(\R^{n})$, then so do the $L\varphi_i$.    
\end{theorem}
{\bf Proof} We do not mimick the proof in \cite{Li83}, but give a somewhat different, and short, argument. We have 
$$
   L\varphi_i(y) = \sqrt{v(y)} \varphi_i \Bigl( \underbrace{\int_{-\infty}^{y_1} (M_1 v)(s) \, ds}_{=:F(y_1)} \Bigr).
$$
Differentiating and using the notation from (\ref{standardmap}) gives 
$$
   \nabla L\varphi_i(y) = \nabla\sqrt{v(y)} \varphi_i(F(y_1)) + \sqrt{v(y)} \varphi'_i(F(y_1)) M_1 v(y_1) e_1,
$$
where $e_1$ is the unit vector in the coordinate direction $y_1$. The first term is in $L^2(\R^n)$ by the boundedness of $\varphi_i$, and the second term
is in $L^2(\R^n)$ by the boundedness of $\varphi'_i$ provided we can show that $M_1 v$ belongs to $L^\infty(\R)$. 
But the latter follows from rewriting $\sqrt{v(y_1,y_2,..,y_n)}^2$ as the integral  
of its derivative with respect to the first variable over $(-\infty,y_1)$, which yields  
$$
   (M_1 v)(y_1) = \int_{\R^{n-1}} \sqrt{v(y_1,..,y_n)}^2 dy_2\cdots dy_n = \int_{\R^n} 1_{y_1> s} \sqrt{v}(s,y_2,..,y_n) 
   \frac{\partial\sqrt{v}}{\partial s}(s,y_2,..,y_n) \, ds \, dy_2\cdots dy_n,
$$
and applying the Cauchy-Schwarz inequality. 

\begin{small}

\end{small}

\end{document}